\documentclass[preprint,12pt]{elsarticle}
\usepackage{amsmath}
\usepackage{graphicx}
\usepackage{dcolumn}
\usepackage{bm}
\usepackage{caption}
\usepackage{float}
\usepackage{xcolor}
 \RequirePackage{mathptmx} 
\journal{Chinese Journal of Physics}

\begin{document}

\begin{frontmatter}

\title{Nuclear structure properties and weak interaction rates of even-even Fe isotopes}
\author{{Jameel-Un Nabi$^{1}$, Mahmut B\"{o}y\"{u}kata$^{2}$, Asim Ullah$^{3}$ and Muhammad Riaz$^{4}$}}

\address{$^{1}$University of Wah, Quaid Avenue, Wah Cantt 47040, Punjab, Pakistan.}
\address{$^{2}$Department of Physics, Faculty of Engineering and Natural Sciences, University of K\i r\i kkale, 71450, K\i r\i kkale, T\"{u}rkiye}
\address{$^{3}$Department of Physics, University of Swabi, Swabi 23561, KP, Pakistan}
\address{$^{4}$Department of Physics, University of Jhang, Jhang, 35200, Punjab, Pakistan.}
\ead{ jameel@uow.edu.pk, boyukata@kku.edu.tr, asimullah844@gmail.com, and mriazgiki@gmail.com}

\begin{abstract}
Nuclear structure properties and weak interaction rates of neutron-rich even-even iron (Fe) isotopes (A = 50 -- 70) are investigated using the Interacting Boson Model-1 (IBM-1) and the proton-neutron Quasiparticle Random Phase Approximation (pn-QRPA) model. The IBM-1 is used for the calculation of energy levels and the B(E2) values of neutron-rich Fe isotopes. Later their geometry was predicted within the potential energy formalism of the IBM-1 model. Weak interaction rates on neutron-rich nuclei are needed for the modeling and simulation of presupernova evolution of massive stars. In the current study, we investigate the possible effect of nuclear deformation on stellar rates of even-even Fe isotopes. The pn-QRPA model is applied to calculate the weak interaction rates of selected Fe isotopes using three different values of deformation parameter. It is noted that, in general, bigger deformation values led to smaller total strength and larger centroid values of the resulting Gamow-Teller distributions. This later translated to smaller computed weak interaction rates. The current finding warrants further investigation
before it may be generalized.The reported stellar rates are up to 4 orders of magnitude smaller than previous calculations and may bear astrophysical significance.

\end{abstract}
\end{frontmatter}
\section{Introduction}
\label{intro}
{The reactions, mediated by weak interaction, play crucial role during the evolutionary phases of massive stars \citep{Bet90,Lan03}. The conversion of proton into neutron via the weak interaction initiates the hydrogen burning in the stellar core of the massive star. The core of the star then goes through He, C, Ne, O and Si burning phases. After the Si burning phase, the core of the star has developed into Fe-peak nuclei. When the mass of the Fe core surpasses the Chandrasekhar mass limit ($\sim$ 1.5 M$_\odot$), the electron degeneracy pressure is no longer capable of supporting the gravitational contraction, and the core begins to collapse. The core-collapse dynamics is sensitive to the temporal variation of the lepton-to-baryon fraction (Y$_e$) and entropy of the core \citep{Bet79}. These quantities are primarily determined by the weak interaction rates (mainly electron capture (\textit{ec}) and $\beta$-decay (\textit{bd})). The \textit{ec} reduces the lepton-to-baryon fraction, which lowers the electron degeneracy pressure while the \textit{bd} counters this effect. At the same time, the \textit{ec} and \textit{bd} rates produce neutrinos and anti-neutrinos, that escape the star with core densities  $\rho$ $\leq$ 10$^{11}$ gcm$^{-3}$, carrying energy and entropy away from the core and may expedite the collapse process \citep{Heg01,Jan07}. These neutrinos and anti-neutrinos are of vital importance during  the stellar evolutionary process. They provide information on neutronization caused by \textit{ec}, the infall phase, shock wave formation and propagation, and the cooling phase of the stellar core.}
on the details of the ground and excited states GT strength distributions. During later phases, once the chemical potential exceeds the Q-value,  the centroid and total GT strength values become the key factors in determining the stellar rates. A reliable calculation, preferably microscopic in nature, of ground and excited states
GT distributions is in order for computation of weak interaction rates and $\beta$-decay half-lives.

Numerous attempts have been made in the past to calculate weak interaction rates in stellar matter. Noticeable mentions would include investigations by Fuller et al. \citep{Ful80,Ful82,Ful82a,Ful85}, Aufderheide et al. \citep{Auf90,Auf94}, Langanke et al. \citep{Lan98,Lan00}, Juodagalvis et al. \citep{Jug10} and
Pruet et al. \citep{Pru03}. Fine-grid calculations of stellar rates can be found in Refs. \citep{Nab99,nab04}.
Weak interaction rates on the \textit{fp}-shell nuclei have a significant impact on the presupernova evolution of heavy-mass stars. Electron capture
cross sections and \textit{ec} rates for various \textit{fp}-shell nuclei were recently
computed using the pn-QRPA model \citep{nab19}.  Various experimental studies were performed for medium-mass nuclei in the
A$\sim$50 region \citep{NNDC22}. More recently, Nabi et al. \citep{nab20} investigated the
nuclear structure properties and impact of nuclear deformation on the calculated \textit{ec}
of even-even chromium isotopes located in the same region of nuclear chart.

Isotopes of Fe  play a  crucial role in the evolution of heavy mass stars. During the last decade, even-even Fe isotopes were actively
investigated \citep{Bai11, Pritychenko12, Pritychenko16, Kaneko14, Coraggio14, Kotila14, Jiao14, Karampagia16} using varying nuclear  models  including the shell model (SM),  the quasiparticle random phase approximation (QRPA) and
the interacting boson model-1 (IBM-1). Fe isotopes, selected for the current investigation,
are amongst the most relevant weak interaction nuclei according to the survey performed by Aufderheide et al. \citep{Auf94}. Heger et al. \citep{Heg01}
rated $^{54,55,56}$Fe amongst top seven (top three) nuclei considered to be most important for decreasing $Y_e$ in 15$M_\odot$ and 40$M_\odot$ (25$M_\odot$) stars. A more recent  simulation study of presupernova evolution of massive stars~\citep{Nab21} highlighted the importance of Fe isotopes. These investigations motivated us to study nuclear properties and stellar weak rates of even-even neutron-rich Fe isotopes.

In this work, we focused on the calculation of nuclear structure properties and stellar rates of even-even $^{50-70}$Fe isotopes
within the framework of IBM-1 and pn-QRPA models. The energy levels and B(E2) values were calculated within the IBM-1
by fitting the parameters of the model Hamiltonian. For the fitting process, the energy ratios (R$_{4/2}$) for the Fe isotope
were analyzed along the isotopic chain to see their behavior. Moreover, their geometric shapes were predicted by plotting the potential energy surface as a function of the deformation parameters. The pn-QRPA model was later applied to calculate weak interaction rates of Fe isotopes as a function of three different deformation values. These included \textit{ec}, \textit{bd}, positron capture (\textit{pc}) and positron decay (\textit{pd}) rates. We, however, report only the dominant rates for each Fe isotope. \\
The paper is structured as follows. Both IBM-1 and pn-QRPA models are briefly explained in Section~2. The results are discussed in Section~3. Summary and conclusion remarks are
given in the final section.
\section{Nuclear Models}
\subsection{Interacting Boson Model-1 (IBM-1)}
\label{ss_ibm}
The IBM-1 was formulated in the 1970s by Iachello and Arima~\citep{Iachello87}
to describe the even-even nucleus by taking into account the given nucleus
as an inert core combined with the bosons. This model is based on the shell
and geometrical models. The IBM-1 has deep relations
with the group theory since it was established on the unitary group $U(6)$
having three possible subgroups "U(5), O(6), SU(3)" named as the
dynamical symmetries. Each symmetry has a relation with the
geometry of the nucleus; the U(5) defines the spherical (vibrational) nucleus,
the SU(3) corresponds to the axially deformed (prolate/oblate) nucleus and
the O(6) describes the asymmetric deformed ($\gamma$-unstable) nucleus.
Model Hamiltonian can be formulated for each
symmetry or by mixing them according to the structure of the given isotope.
For the present application, the following Hamiltonian was used:
\begin{equation}
	\hat H= \epsilon_d\,\hat n_d + a_0 \,\hat P_+\cdot\hat P_- + a_1 \,\hat Q\cdot\hat Q + a_2\,\hat L\cdot\hat L. \label{ham}
\end{equation}
Here, $\hat n_d$ is the boson number operator, $\hat P$ is the pairing operator,
$\hat L$ the angular momentum operator, and $\hat Q$ is the quadrupole operator~\citep{Iachello87}.
The constants of Hamiltonian ($\epsilon_d$, $a_0$, $a_1$ and $a_2$) are free parameters
fitted from the recent experimental data obtained from the National
Nuclear Data Center (NNDC)~\citep{NNDC22}.

These operators were defined depending on combinations of the bosonic creation and annihilation
operators $s$, $s^{\dag}$, $d$, $d^{\dag}$ as follows:
\begin{equation} \label{opr}
	\begin{aligned}
		\hat n_d = \sqrt{5}[d^\dag\times\tilde d]^{(0)}  ,  \hat L = \sqrt{10}[d^\dag\times\tilde d]^{(0)} \nonumber\\
		\hat P_+ = [s^\dag\times s^\dag+ \sqrt{5} (d^\dag\times d^\dag) d]^{(0)}  ,  \hat P_- = (\hat P_+)^\dag \nonumber\\
		\hat Q=[d^\dag\times\tilde s+s^\dag\times\tilde d]^{(2)}+\overline{\chi}[d^\dag\times\tilde d]^{(2)},
	\end{aligned}
\end{equation}
where $\overline{\chi}$ in the quadrupole operator is another free parameter fitted using the same procedure as others.

The energy levels of Fe isotopes were calculated using the Hamiltonian in Eq.~(\ref{ham}). The B(E2) values
were calculated  using the quadrupole transition operator:
\begin{equation} \label{be2}
	\hat T (E2) = e_b \cdot \hat Q \nonumber\\,
\end{equation}
where the constant $e_b$ is the boson effective charge
used as a free parameter for the calculation.

The prediction of the geometry of the nucleus is of paramount importance for the investigation of the nuclear structure.
The IBM-1 model is based on the geometric collective model~\citep{Bohr98}. The geometric
shape of the given nuclei may be predicted by plotting the potential energy surface as a function of the deformation
parameters. The potential energy surface in the classical limit ~\citep{Dieperink80,Dieperink80b,Ginocchio80,Isacker81} can be written as follows:

\begin{eqnarray}\label{pes}
	\resizebox{0.019\hsize}{!}
	V(\beta,\gamma)=\epsilon_d \frac {N \beta^2}{1+\beta^2}
	+ a_0 \frac{ N (N-1) }{4} \left(\frac{1-\beta^2}{1+\beta^2}\right)^2
	+ a_2 \frac{6 N \beta^2}{1+\beta^2}
	+ a_1 \frac{ N }{1+\beta^2} ~~~~~~~~~~~~~~~~~~~~~~~~~~~~~~~~~~~~~ \nonumber\\
	\times \left[ 5 + (1 + \overline{\chi}^2) \beta^2 +
	\frac{(N-1)\left(\frac{2\overline{\chi}^2\beta^2}{7}-4\sqrt{\frac{2}{7}}\overline{\chi}\beta\cos3\gamma+4\right)\beta^2}{1+\beta^2}\right].
\end{eqnarray}
The deformation parameters ($\beta$ and $\gamma$) have a similar role as in the geometric collective model~\citep{Bohr98}. The parameter
$\beta$ measures the axial drift from the spherical to the deformed region whereas the $\gamma$ determines the shift from the prolate to oblate side in the deformed region. These parameters can be calculated by minimizing the potential energy surface given in Eq.~(\ref{pes}). The $\beta$ deformation can also be determined from the reduced electric quadrupole transition rate by using the following formula~\citep{ram01}:
\begin{equation}
\beta=\left(\frac{4\pi}{3 Z R^2_0}\right) \left[\frac{B(E2)\uparrow}{e^2}\right]^{1/2}, \label{def2}
\end{equation}
where $B(E2)\uparrow$  is  the reduced electric quadrupole transition rate from the ground to the first excited state [B(E2:$0^+_{1}\rightarrow2^+_{1}$)],
$Z$ is the number of protons, and R$_0$ is the nuclear radius given by:
\begin{equation}
R^2_0=(1.2\times 10^{-13} R^{1/3} cm)^2=0.0144 A^{2/3}b. \label{R0}
\end{equation}
The $B(E2)\uparrow$ may be calculated  from the $B(E2)\downarrow$ value as follows:
\begin{equation}
B(E2)\uparrow =\left(\frac{2J_f + 1}{2J_i + 1}\right) B(E2)\downarrow, \label{BE2up}
\end{equation}
where $B(E2)\downarrow$  is the transition from the first excited to the ground state [B(E2:$2^+_{1}\rightarrow0^+_{1}$)].
The experimental B(E2) values can be obtained from the Evaluated Nuclear Structure Data File (ENSDF) of the NNDC~\citep{NNDC22}.

\subsection{The proton-neutron quasiparticle random phase approximation (pn-QRPA) model}
The pn-QRPA Hamiltonian employed to calculate GT strength distributions and associated weak interaction rates was of the form:
\begin{equation} \label{H}
H^{QRPA} = H^{sp} + V^{ph}_{GT} + V^{pp}_{GT} + V^{pair},
\end{equation}
where $H^{sp}$ stands for the single-particle Hamiltonian. The residual Hamiltonian consisted of three terms. $V_{GT}^{ph}$
and $V_{GT}^{pp}$ represent the particle-hole and particle-particle GT
forces, respectively. The last term $V^{pair}$ denotes the pairing
force and was determined using the BCS approximation. 
The constant pairing force of strength G ($G_p$ and $G_n$ for protons and neutrons, respectively), is given by
	\begin{eqnarray}\label{pr}
		V_{pair}=-G\sum_{jmj^{'}m^{'}}(-1)^{l+j-m}c^{\dagger}_{jm}c^{\dagger}_{j-m}\\ \nonumber
		(-1)^{l^{'}+j^{'}-m^{'}} c_{j^{'}-m^{'}}c_{j^{'}m^{'}},
	\end{eqnarray}
where	$c^{\dagger}$ and $c$ represents the particle creation and annihilation operators in the spherical basis.   The summation over $m$ and $m^{'}$ was limited to $m$, $m^{'}$ $>$ 0, where $m$ is the projection of angular momentum on the symmetry axis and the remaining symbols have their usual meaning. The separable $ph$ GT force can be expressed as
\begin{equation}\label{ph}
	V^{ph}= +2\chi\sum^{1}_{\mu= -1}(-1)^{\mu}Y_{\mu}Y^{\dagger}_{-\mu},\\
\end{equation}
with
\begin{equation}\label{y}
	Y_{\mu}= \sum_{j_{p}j_{n}}<j_{p}m_{p}\mid
	t_- ~\sigma_{\mu}\mid
	j_{n}m_{n}>c^{\dagger}_{j_{p}m_{p}}c_{j_{n}m_{n}},
\end{equation}
where  $\mu=m_{p}-m_{n}$ = 0, $\pm$1.
The separable $pp$ GT force was approximated using
\begin{equation}\label{pp}
	V^{pp}= -2\kappa\sum^{1}_{\mu=
		-1}(-1)^{\mu}P^{\dagger}_{\mu}P_{-\mu},
\end{equation}
with
\begin{eqnarray}\label{p}
	P^{\dagger}_{\mu}= \sum_{j_{p}j_{n}}<j_{n}m_{n}\mid
	(t_- \sigma_{\mu})^{\dagger}\mid
	j_{p}m_{p}>\times \nonumber\\
	(-1)^{l_{n}+j_{n}-m_{n}}c^{\dagger}_{j_{p}m_{p}}c^{\dagger}_{j_{n}-m_{n}},
\end{eqnarray}
where $\sigma_{\mu}$ denotes the spherical components of the spin operator and $t_{-}$ is the isospin lowering operator. The different signs in \textit{ph} and \textit{pp} force reveal the opposite nature of these interactions (\textit{pp} force is attractive while the \textit{ph} force is repulsive).
 The single-particle energies and wavefunctions were determined using the Nilsson model \citep{Nil55}. The Nilsson models require nuclear deformation ($\beta$) as one of its model parameters. The $pp$ GT force was determined using interaction strength parameter ($\kappa$) whereas the $ph$ GT force was determined using interaction strength parameter ($\chi$) in accordance with Ref.~\citep{Hom96}, based on a $1/A^{0.7}$ relationship. The oscillator constant for nucleons was computed using $\hbar\omega=\left(45A^{-1/3}-25A^{-2/3}\right)$ MeV \citep{Blo68} (same for protons and neutrons). The Nilsson-potential
parameters were adopted from Ref.~ \citep{ragnarson1984}.
$Q$-values were taken from the recent mass compilation of Ref. \citep{Aud21}. The pairing gaps for proton and neutron were computed using separation energies of proton (S$_p$) and neutron (S$_n$), respectively, using:
\begin{eqnarray}
\bigtriangleup_{pp} =\frac{1}{4}(-1)^{Z+1}[S_p(Z+1, A+1)-2S_p(Z, A)+S_p(Z-1, A-1)]
\end{eqnarray}
\begin{eqnarray}
\bigtriangleup_{nn} =\frac{1}{4}(-1)^{A-Z+1}[S_n(Z, A+1)- 2S_n(Z, A) + S_n(Z, A-1)]
\end{eqnarray}
For solution of the Hamiltonian we refer to~\citep{Hom96}.

The weak interaction rate from the $\mathit{i}^{th}$ state of parent nucleus to the $\mathit{j}^{th}$ state of daughter nucleus is given by:
\begin{equation}
\lambda_{ij} =ln2
\frac{\phi_{ij}(\rho, T, E_{f})}{(ft)_{ij}},
\label{d_rate}
\end{equation}
where $\phi_{ij}$ are the phase space integrals over total energy. The symbols $\rho$, $T$ and $E_f$ denote the density, temperature and Fermi energy, respectively. The term $(ft)_{ij}$ is connected with the reduced transition
probability B$_{ij}$ of nuclear transitions by:
\begin{equation}\label{ft} (ft)_{ij}=F/B_{ij}.
\end{equation}
where F is a physical constant, given by:
\begin{equation}
F=\frac{2ln2\hbar^{7}\pi^{3}}{g_{V}^{2}m_{e}^{5}c^{4}} \nonumber,
\end{equation}
and B$_{ij}$ is determined using:
\begin{equation}
B_{ij}=B(F)_{ij}+(g_{A}/g_{V})^2 B(GT)_{ij},
\end{equation}
where the reduced Fermi ($B(F)_{ij}$) and
GT transition probabilities ($B(GT)_{ij}$) were calculated employing the
following relations:
\begin{equation}
B(F)_{ij} = \frac{1}{2J_{i}+1} \mid<j \parallel \sum_{k}t_{\pm}^{k}
\parallel i> \mid ^{2},
\end{equation}
\begin{equation}\label{gt}
B(GT)_{ij} = \frac{1}{2J_{i}+1} \mid <j
\parallel \sum_{k}t_{\pm}^{k}\vec{\sigma}^{k} \parallel i> \mid ^{2}.
\end{equation}
In the last two equations, $J_i$ denotes the total spin of the parent state $|i \rangle$,
$\sigma^{\rightarrow k}$ are the Pauli spin matrices and $t_\pm^k$ refer
to the iso-spin raising and lowering operator. We took $F$ = 6295$s$ and $g_A/g_V$ = -1.254 from Refs. \citep{Nak10}
and ~\citep{hardy09}, respectively.  \\
The phase space integral for the \textit{ec} was determined using:
\begin{equation}\label{pc}
\phi_{ij} = \int_{w_{l}}^{\infty} w (w_{m}+w)^{2}({w^{2}-1})^{\frac{1}{2}} F(+Z,w) G_{-} dw,
\end{equation}
while for \textit{bd} (upper sign) and \textit{pd} (lower sign), the phase space integrals were computed using:
\begin{equation}\label{ps}
\phi_{ij} = \int_{1}^{w_{m}} w (w_{m}-w)^{2}({w^{2}-1})^{\frac{1}{2}} F(\pm Z,w)
(1-G_{\mp}) dw,
\end{equation}
$w$, $w_{m}$ and $w_l$ appearing above denote the total energy of the electron (rest-mass + kinetic), the total \textit{bd} energy and total threshold energy for \textit{ec}, respectively. The Fermi functions ($F(\pm Z, w)$) were computed using the prescription given by Gove and Martin ~\citep{Gov71}. $G_{-}$ and $G_{+}$ stands for the electron and positron distribution functions, respectively. Assuming natural units ($\hbar=m_{e}=c=1$), they are given by
\begin{equation}
	G_- = [exp(\frac{w-E_f}{kT})+1]^{-1},
\end{equation}
\begin{equation}
	G_+ = [exp(\frac{w+2+E_f}{kT})+1]^{-1}.
\end{equation}
The total weak interaction rates were computed using
\begin{equation}\label{ecbd}
\lambda^{ec/bd/pd/pc} =\sum _{ij}P_{i} \lambda^{ec/bd/pd/pc} _{ij}.
\end{equation}
The summation was applied over all the initial and final states until satisfactory convergence in rates was achieved. $P_{i} $ in Eq.~(\ref{ecbd}) denotes the occupation probability of parent excited states and follows the normal Boltzmann distribution.
\section{Results and Discussion}
\begin{figure*}[h!]
	\includegraphics[width=1.2\textwidth]{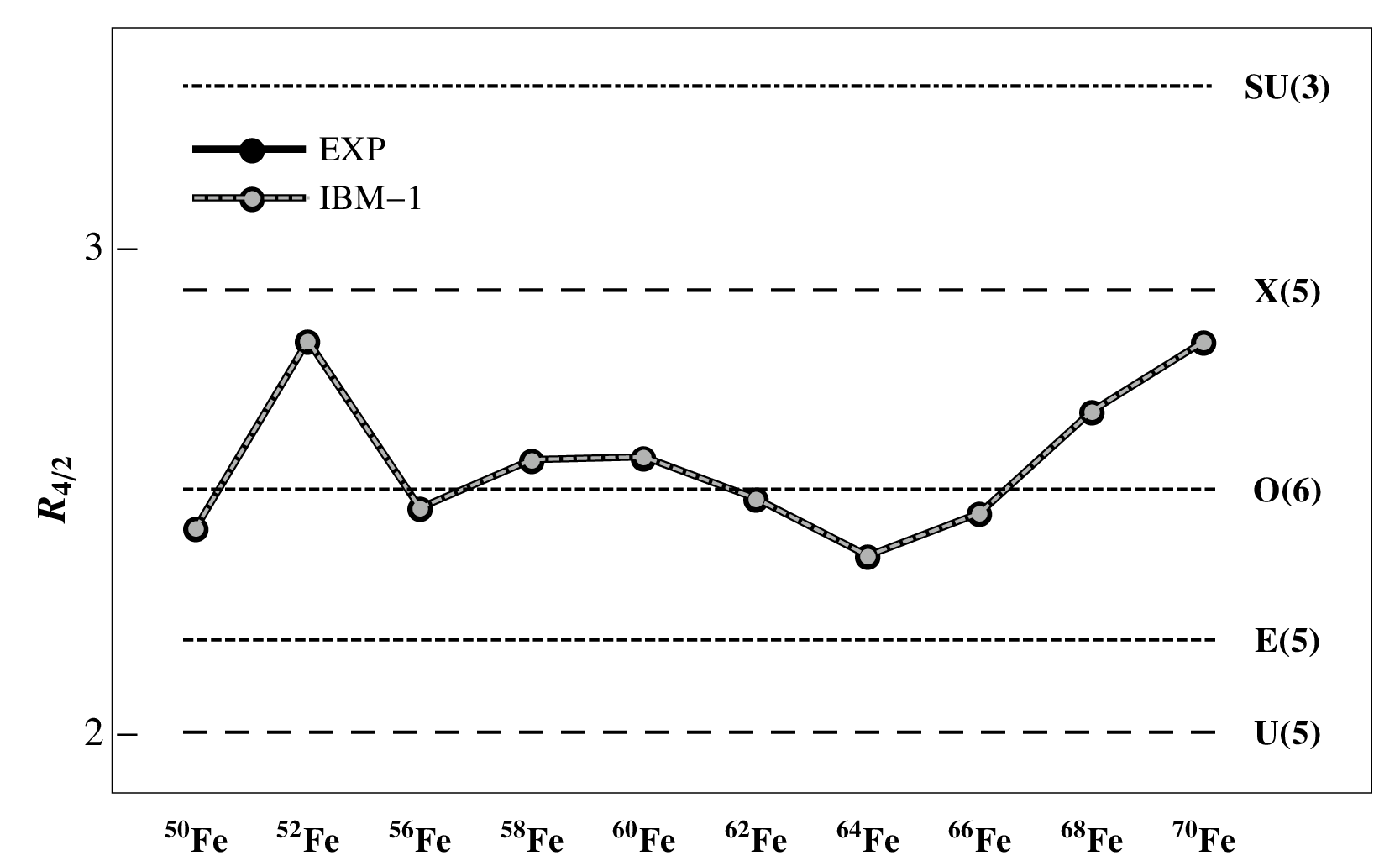}
	\caption{The energy ratios (R$_{4/2}$) of Fe isotopes along the isotopic chain.
	}
	\label{fig:ratio}
\end{figure*}

\begin{figure*}[h!]
	\includegraphics[width=.5\textwidth]{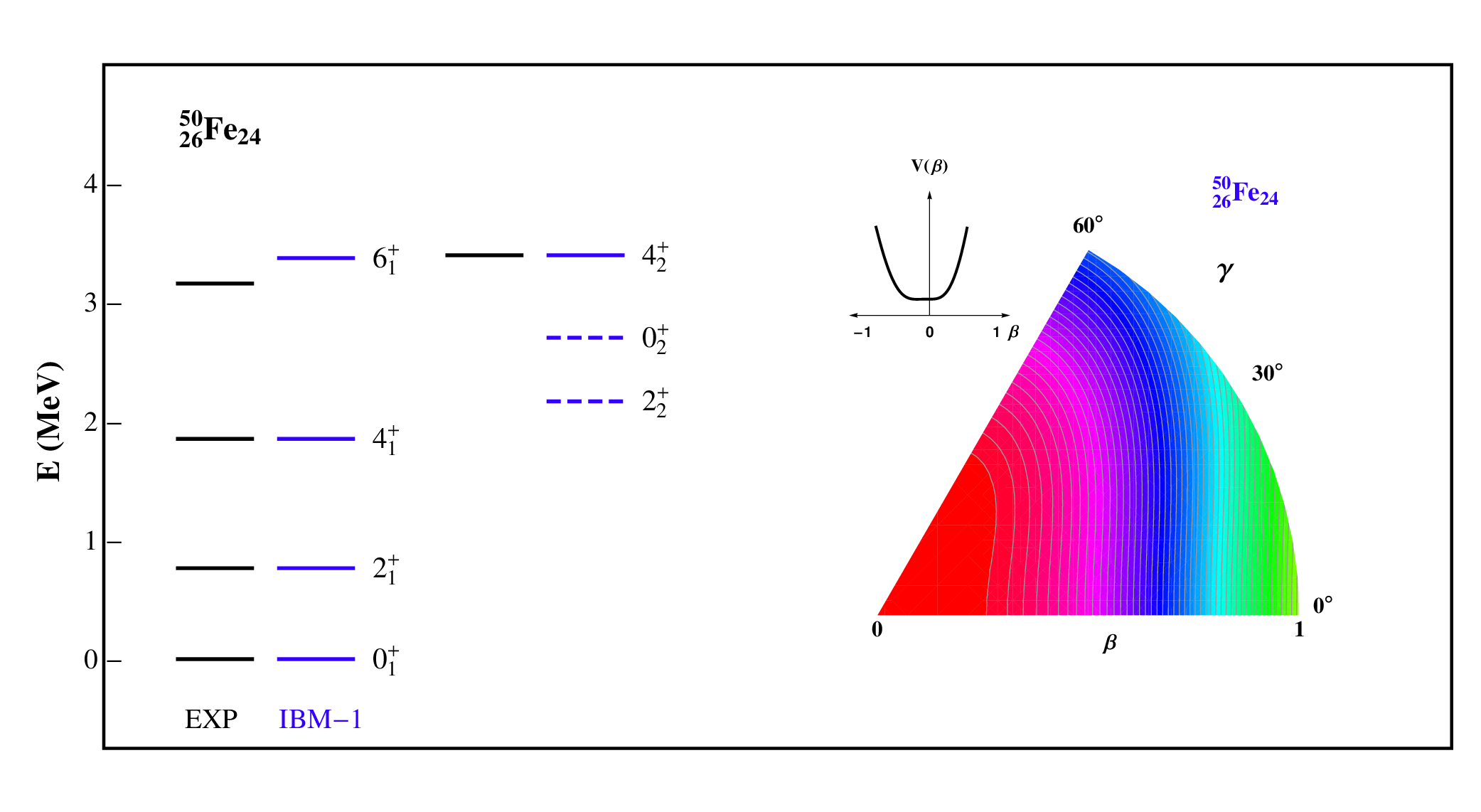}
	\includegraphics[width=.5\textwidth]{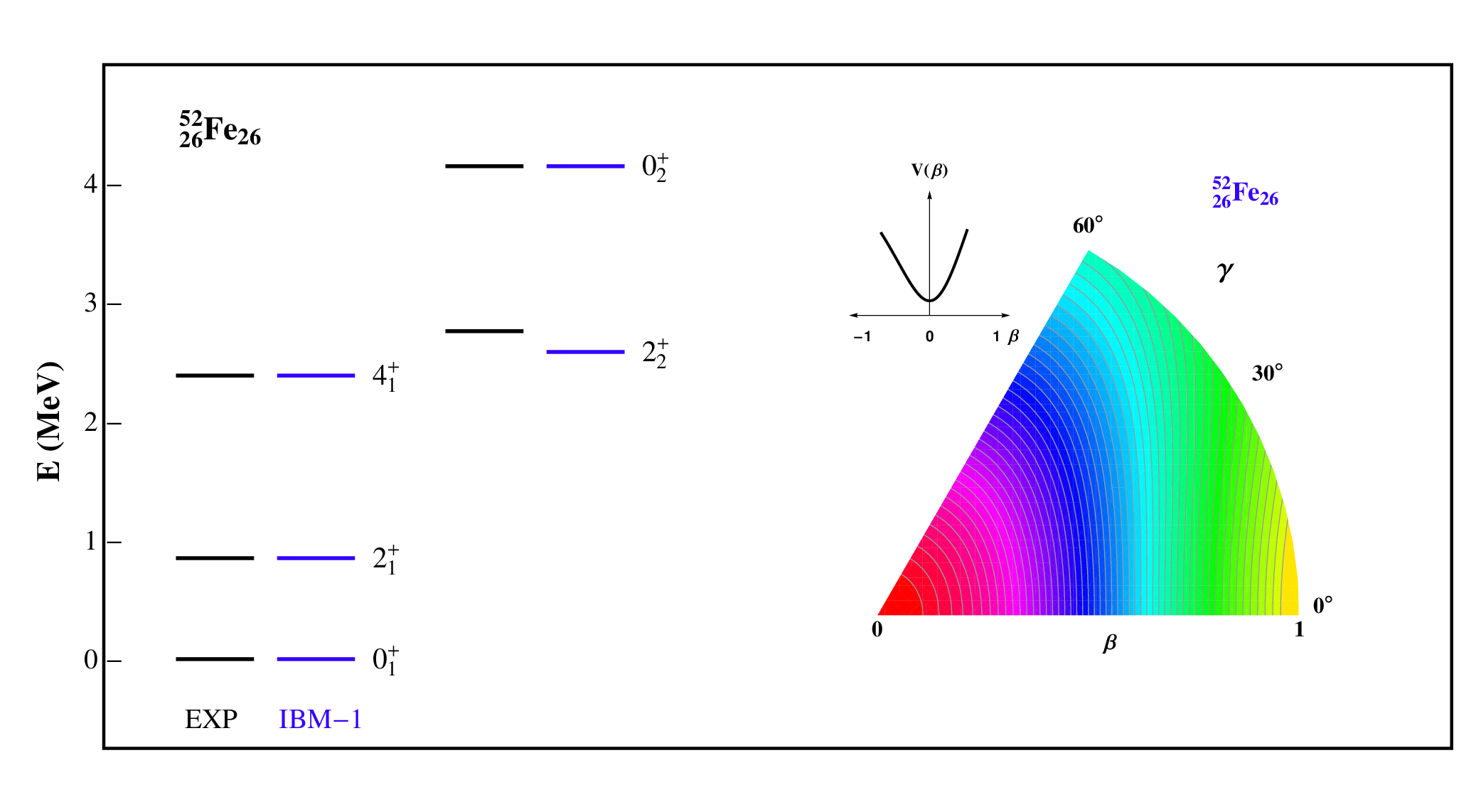}
	\includegraphics[width=.5\textwidth]{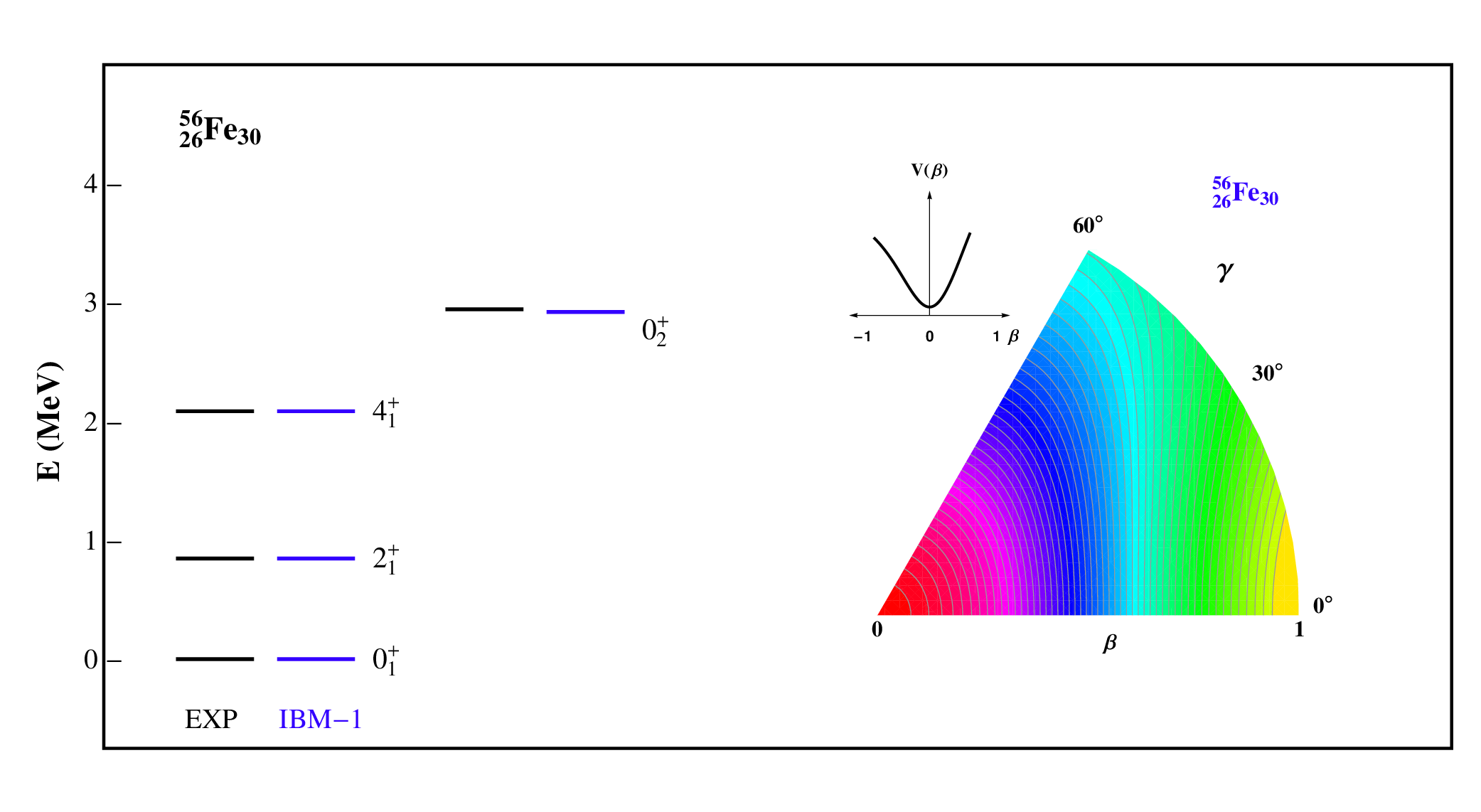}
	\includegraphics[width=.5\textwidth]{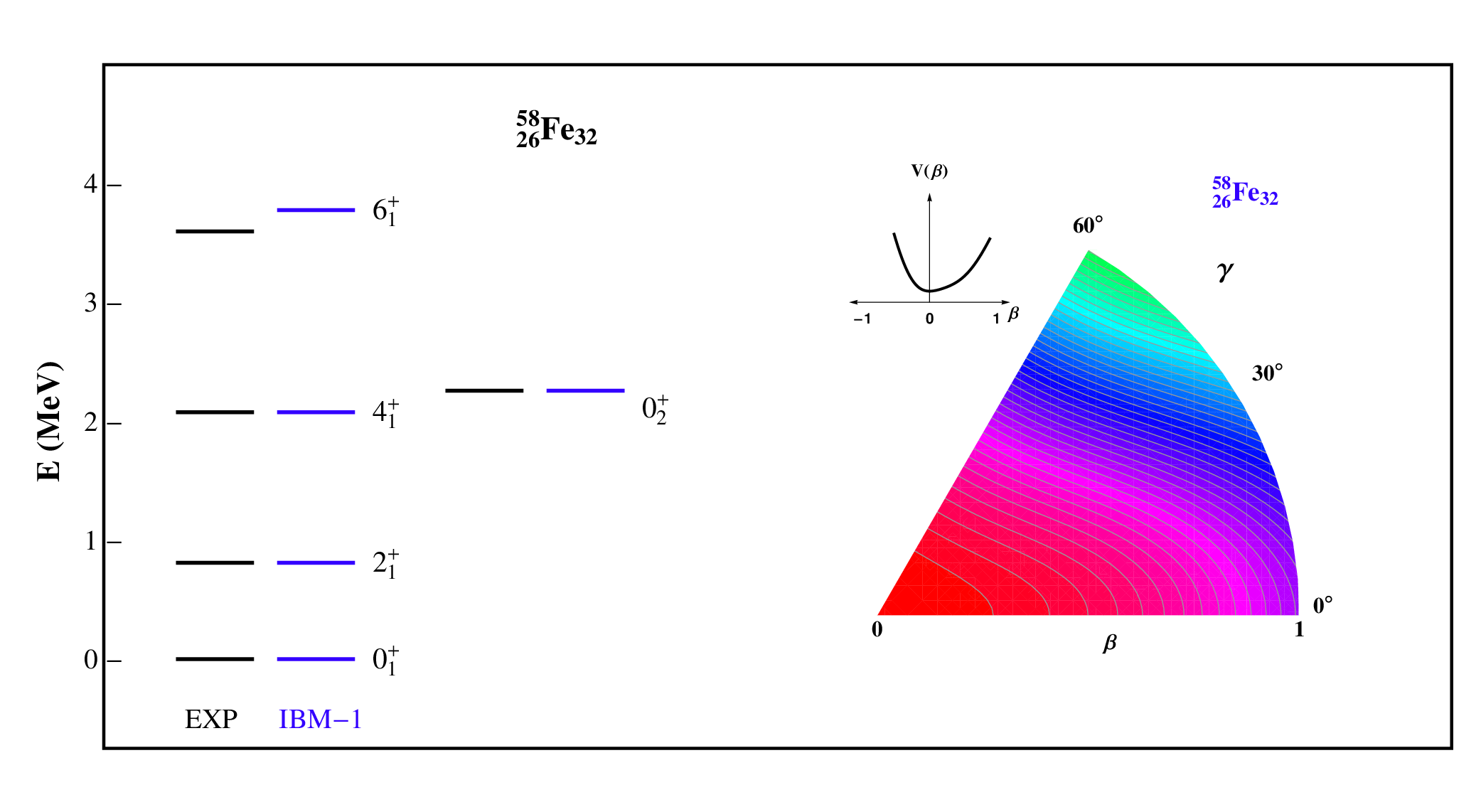}
	\includegraphics[width=.5\textwidth]{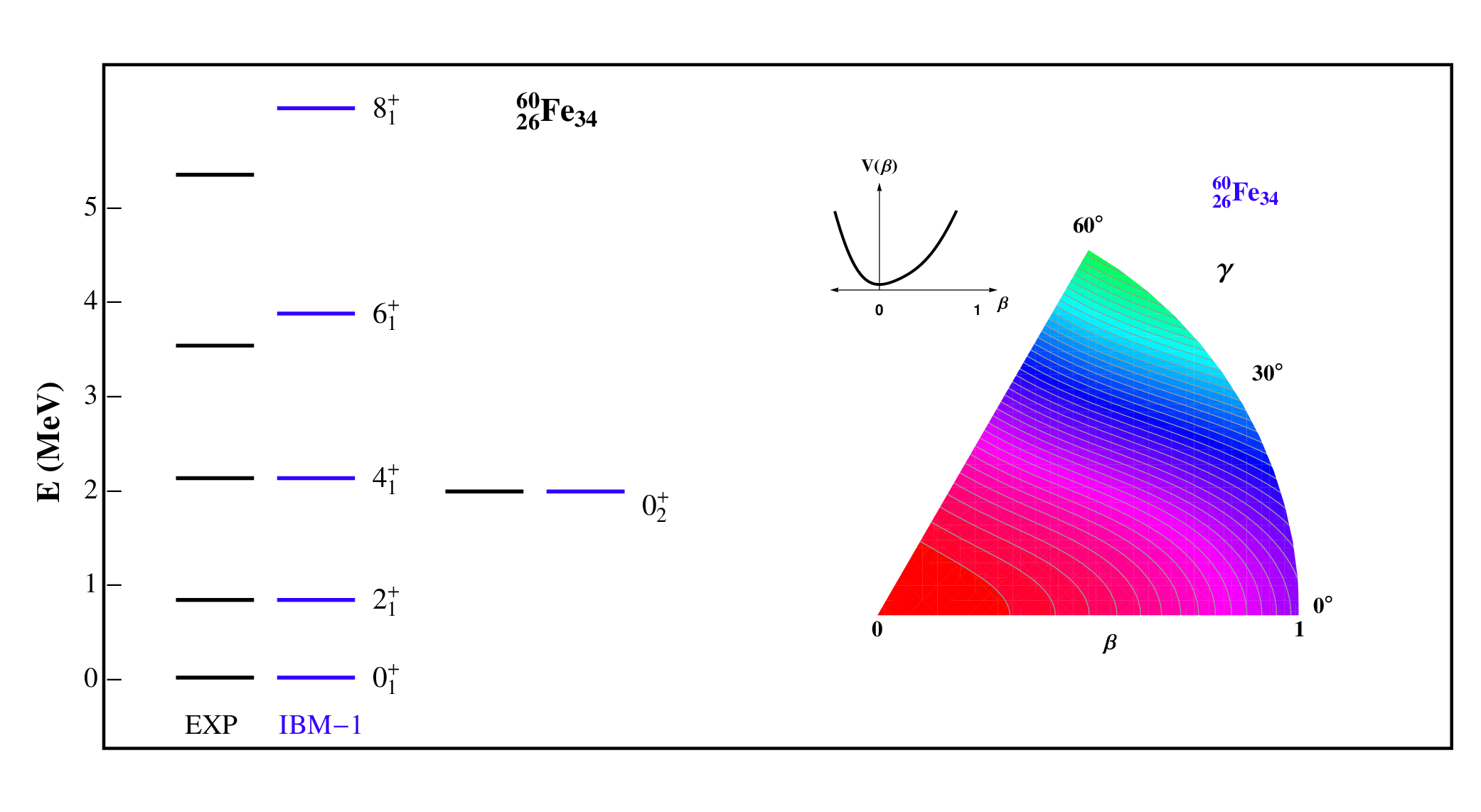}
	\includegraphics[width=.5\textwidth]{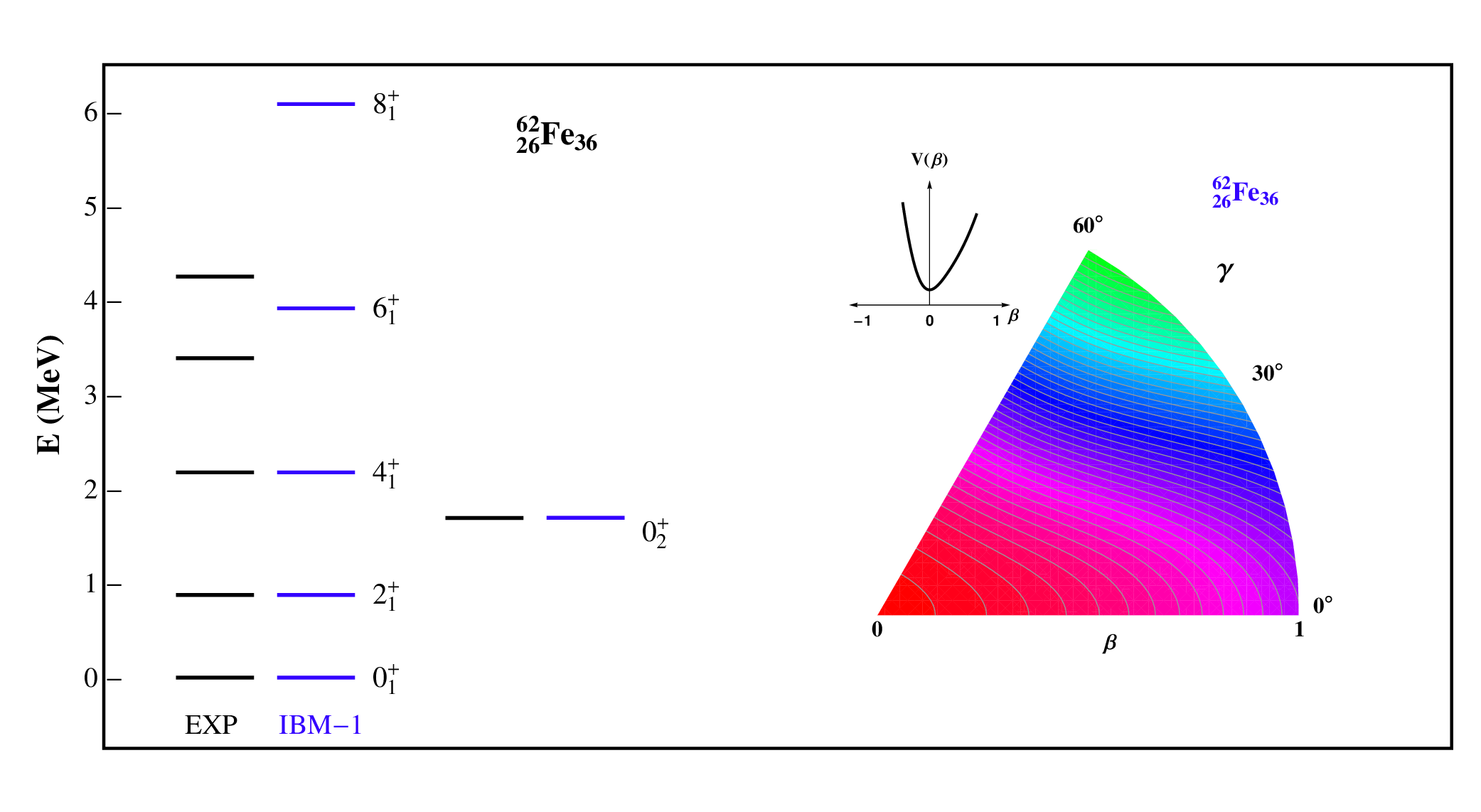}
	\includegraphics[width=.5\textwidth]{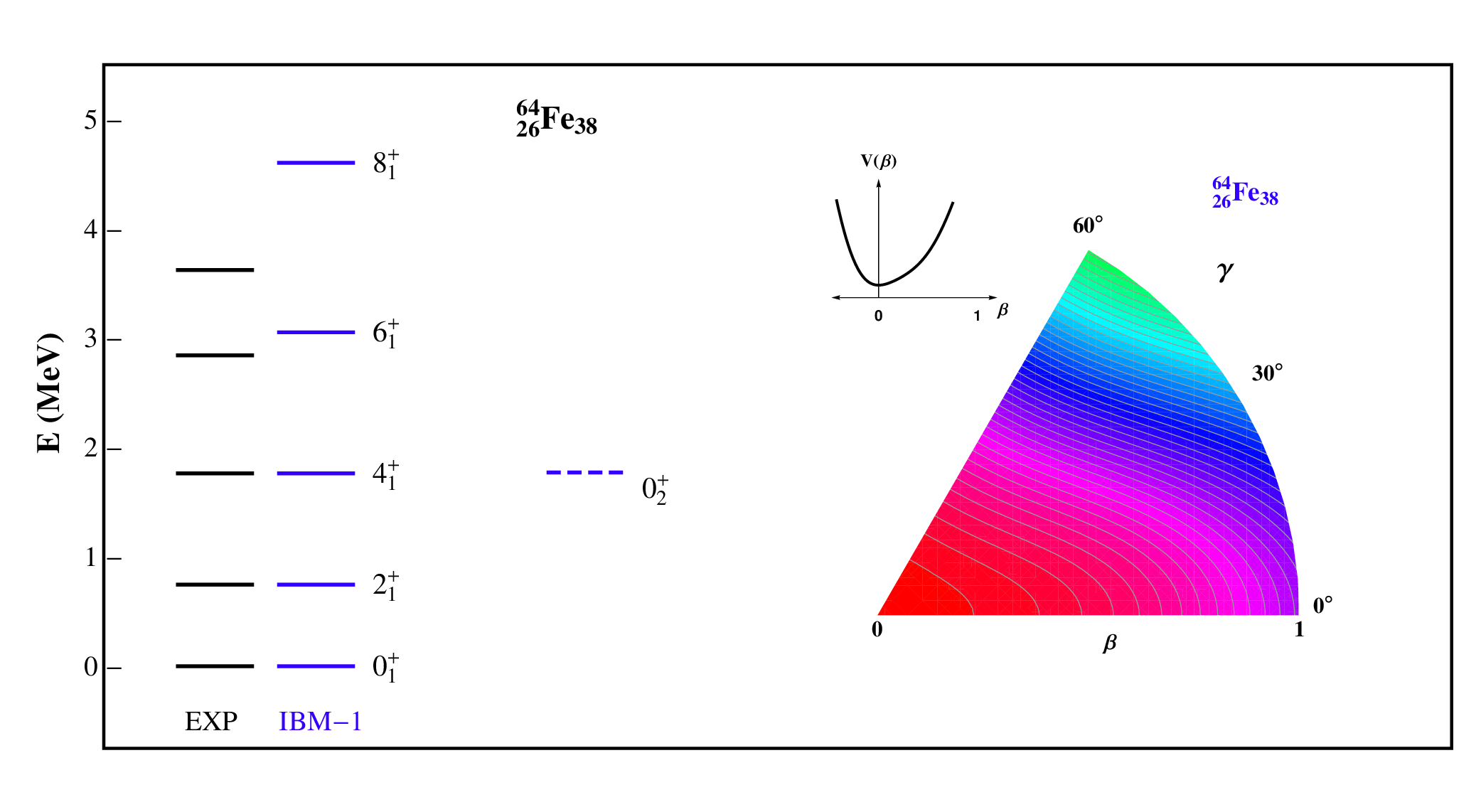}
	\includegraphics[width=.5\textwidth]{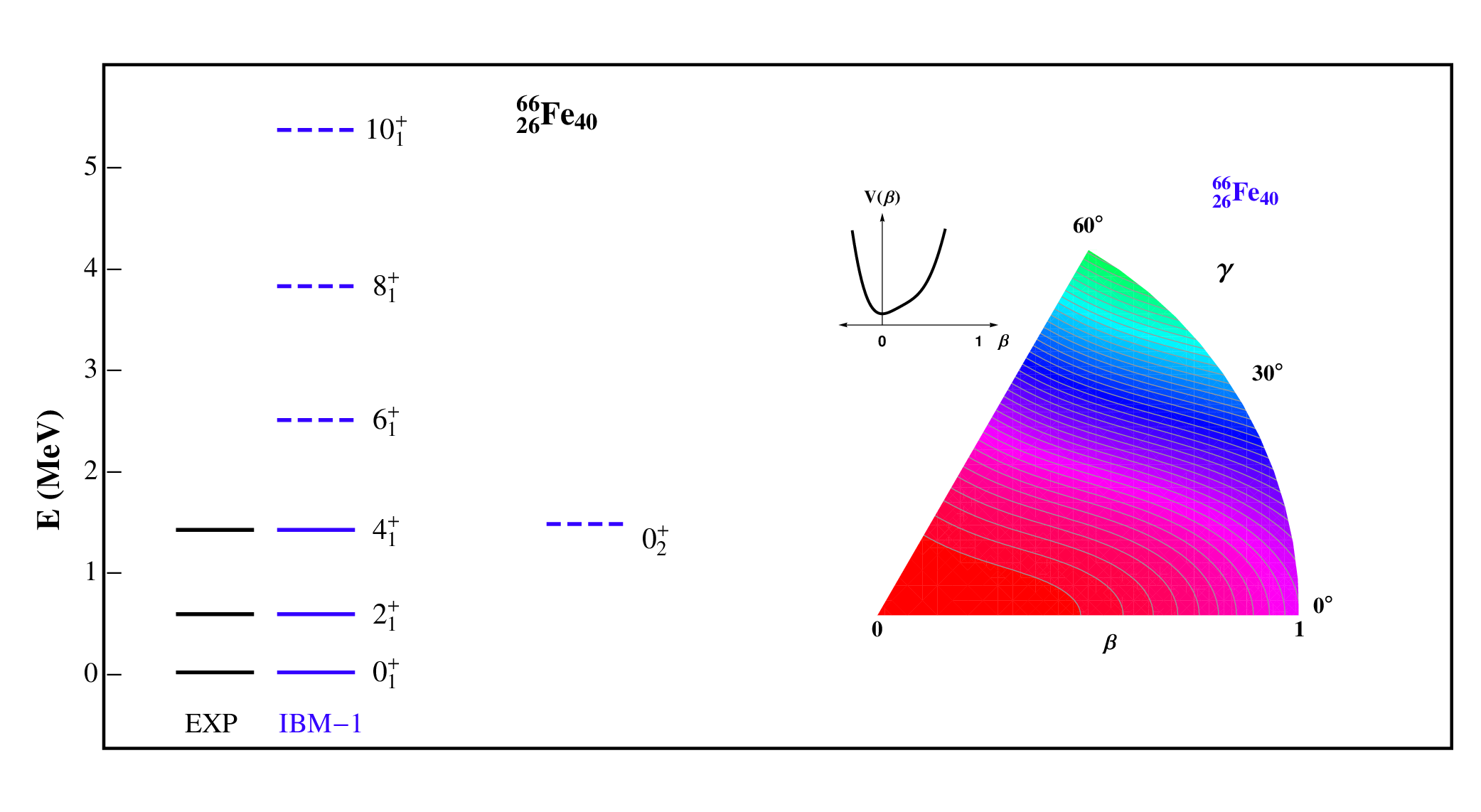}
	\includegraphics[width=.5\textwidth]{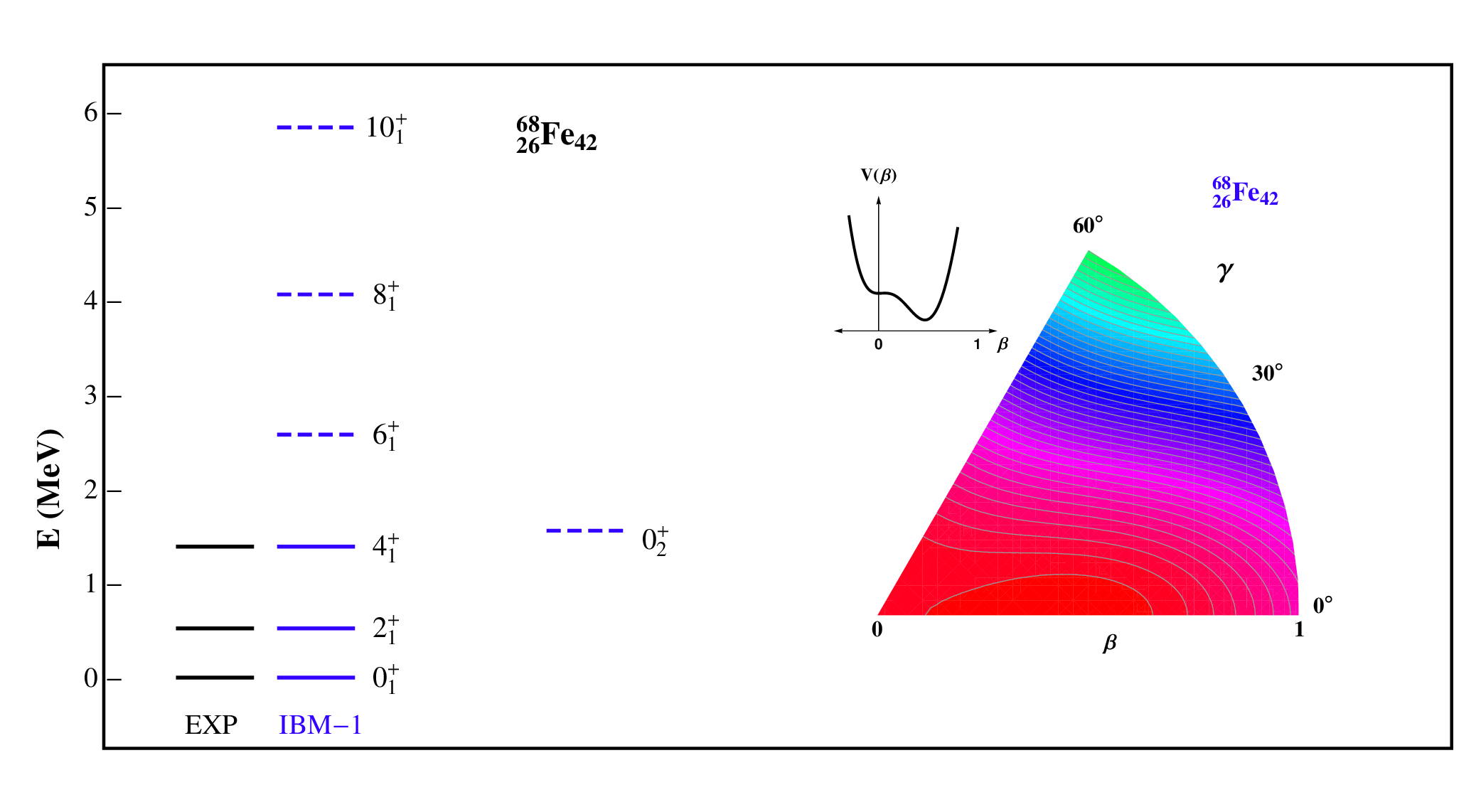}
	\includegraphics[width=.5\textwidth]{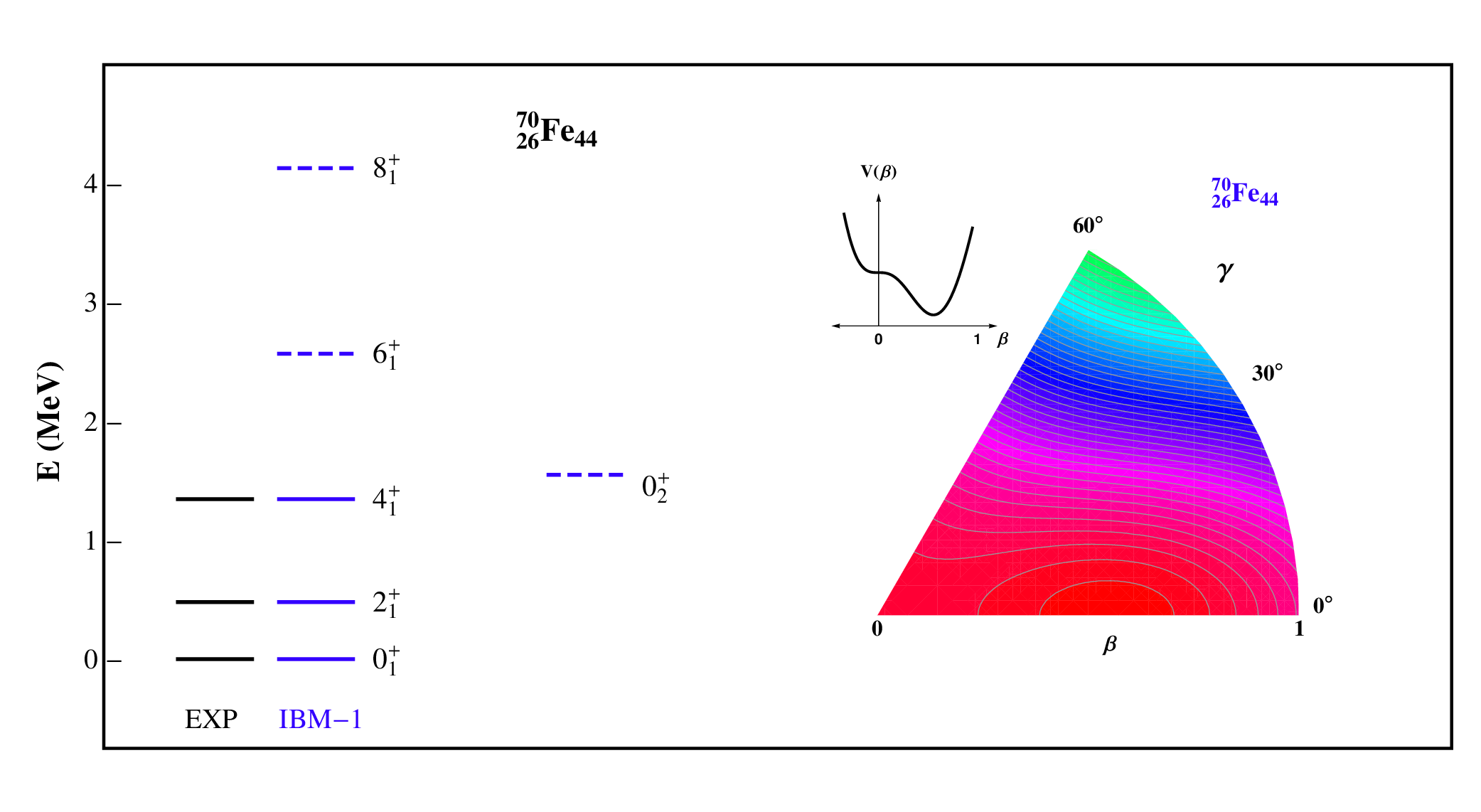}
	\caption{\textcolor{blue}{The experimental~\cite{NNDC22} and calculated energy levels. The right side in each panel depicts the contour plot of the energy surfaces in the ($\beta$, $\gamma$) plane and as a function of $\beta$ for $\gamma$ = 0 for the
	selected Fe isotopes.
	}}
	\label{fig:enpes}
\end{figure*}

\begin{figure}[h!]
	\includegraphics[width=1.2\textwidth]{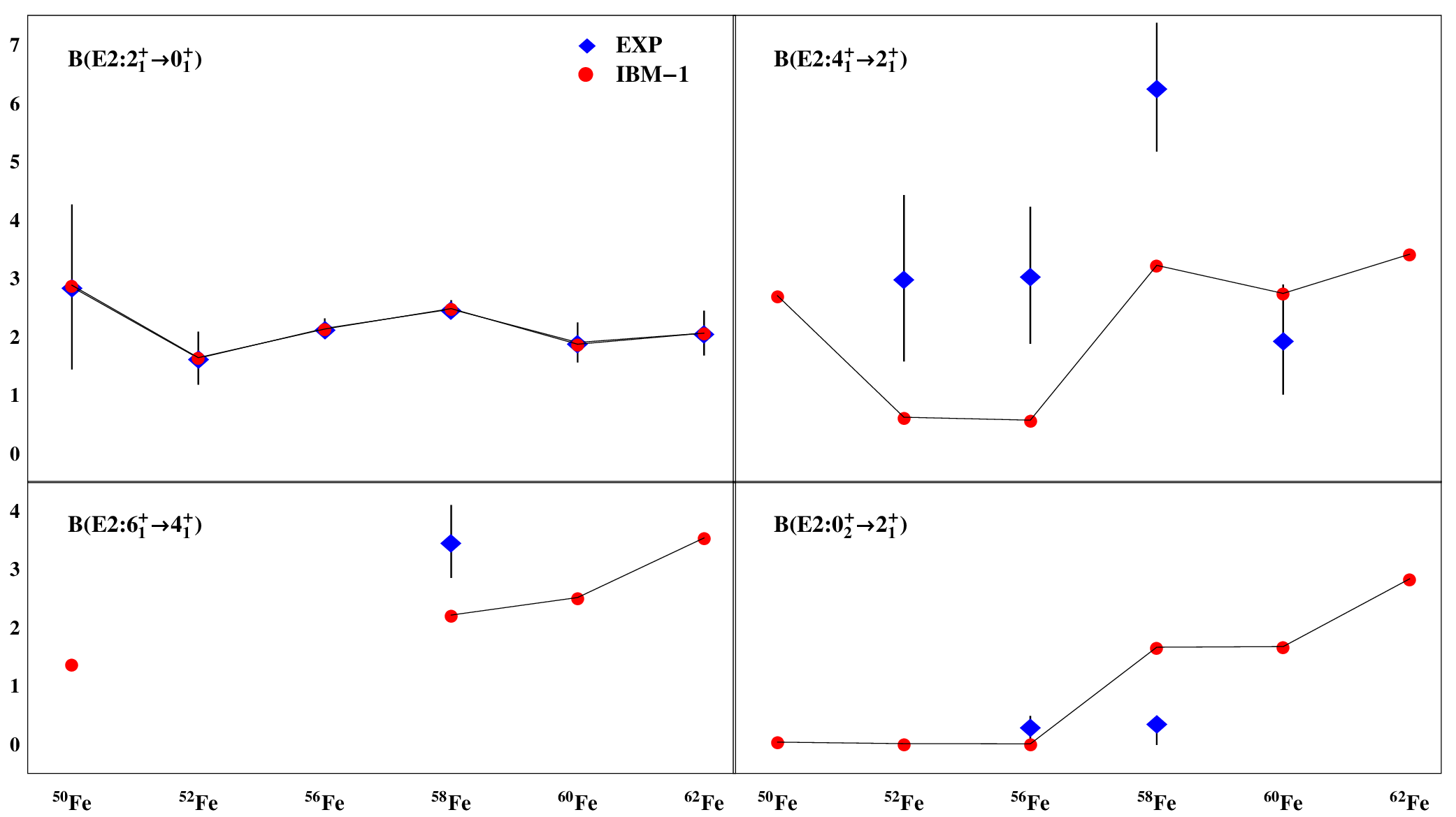}
	\caption{The calculated B(E2)$\downarrow$ values, in units of $10^{-2}$ $e^{2}b^{2}$
		for given even-even Fe isotopes.}
	\label{fig:be2fe}
\end{figure}

In the first phase of the current investigation, we studied the structure of even-even $^{50,52,56-70}$Fe isotopes along the Z=26 chain within the IBM-1 model. Model calculations were performed for the energy levels, B(E2) values and predicted the geometric shapes of these isotopes.
$^{54}$Fe was excluded from the IBM-1 calculation having 28 neutrons (magic number) resulting in its number of neutron bosons as zero. To calculate the energy levels, the Hamiltonian parameters (shown as constants in  Eq.~(\ref{ham})) were fitted by minimizing the root mean square (rms) deviations given by
\begin{eqnarray}\label{rms}
	\sigma&=&\sqrt{\frac{1}{N}\sum_i\left(exp.^i-cal.^i\right)^2}.
\end{eqnarray}
Before the fitting procedure, the energy ratios (R$_{4/2}$) were analyzed along the
isotopic chain of given even-even Fe isotopes to study their behavior.
As seen from Fig.~\ref{fig:ratio}, all isotopes were located around O(6) symmetry.  $^{52,70}$Fe were  closer to X(5) symmetry~\citep{Iachello01}
which is located in between spherical and axially deformed shapes and $^{64}$Fe
is in between O(6) -- E(5)~\citep{Iachello00} symmetries. The X(5) and E(5) are
known as critical point symmetries used to describe critical points in the shape
transition from spherical to axially deformed and from spherical to deformed $\gamma$-unstable
case, respectively~\citep{Iachello01,Iachello00}. For the fitting procedure, the experimental $2^+_1$, $4^+_1$
levels in the ground state band and $0^+_2$ level were taken into account to
fit the Hamiltonian parameters. The rms ($\sigma$) values were obtained as zero for all Fe isotopes.
Fig.~\ref{fig:ratio} shows that the experimental and calculated R$_{4/2}$ ratio are exactly the same.
Later, other experimental levels were added to our calculations and the revised $\sigma$ values with
the set of the fitted Hamiltonian parameters ($n_{d}$, $a_0$, $a_1$, $a_2$)
are listed in Table \ref{par}. The $\overline{\chi}$ was fixed as -1.32 (dimensionless).
This table also includes the fitted boson effective charges ($e_b$) and its rms ($\sigma*$) values along the last two columns.
The calculated energy levels of Fe isotopes including experimental data~\citep{NNDC22}
are exhibited in Fig.~\ref{fig:enpes}. As seen in this figure, the calculations
of energy levels are in good agreement with the measured ones.
Few unknown energy levels were also predicted within the given set of
Hamiltonian parameters in Table \ref{par}.
Fig.~\ref{fig:enpes} contains the contour plots of potential energy surfaces
in $\beta$ and $\gamma$ plane and the energy surfaces as a function of $\beta$ for
$\gamma=0$. The even-even $^{52,56,62}$Fe isotopes are purely
spherical because both their $\beta$ and $\gamma$ values are zero. $^{50}$Fe has flat minima
on the oblate side as seen in the first panel of Fig.~\ref{fig:enpes}. This is a
sign of the critical point for the transition from spherical to axially
deformed (oblate) shape. Although $^{58,60,64,66}$Fe have spherical minima,
they tend to have flat minima in the prolate region. Especially, $^{66}$Fe
isotope is close to the critical point located in between the spherical and  axially
deformed (prolate) shapes. Unlike these isotopes, the $\beta$ values of $^{68,70}$Fe
isotopes are in prolate side and so these two isotopes show prolate behavior.
B(E2)$\downarrow$ values in the low-lying states were also calculated by
using boson effective charges $e_b$ fitted by minimizing rms values given above.
The experimental B(E2:$2^+_{1}\rightarrow0^+_{1}$) were taken into account to
fit $e_b$ and the obtained rms ($\sigma*$) values were listed in the last column of Table \ref{par}
for $^{50,52,56,58,60,62}$Fe isotopes.
As shown in the first panel of Fig.~\ref{fig:be2fe}, the calculated B(E2:$2^+_{1}\rightarrow0^+_{1}$)
values are the same with experimental ones.
The calculated B(E2:$2^+_{1}\rightarrow0^+_{1}$), B(E2:$4^+_{1}\rightarrow2^+_{1}$),
B(E2:$6^+_{1}\rightarrow4^+_{1}$) and B(E2:$0^+_{2}\rightarrow2^+_{1}$) values are
shown in Fig.~\ref{fig:be2fe} for $^{50,52,56,58,60,62}$Fe isotopes. This figures also includes
predicted B(E2) values for unknown ones. The calculated $\beta$ deformation parameters within IBM-1 are
listed in Table~\ref{def}. This table also includes measured deformations
obtained from the experimental reduced electric quadrupole transition rate
by using Eq.~(\ref{def2}) and the FRDM model~\citep{Mol16}.

The pn-QRPA model was used to calculate ground and excited states GT strength distributions and stellar weak rates in the latter phase of the current investigation. In particular, since we used the Nilsson model to calculate the single-particle energies and wavefunctions, nuclear deformation was a free parameter in our pn-QRPA calculations. We investigated the effect of deformation parameter, studied during the first phase of this project, on the calculated weak interaction rates. Three different deformation values were used to compute the weak interaction rates. The IBM-1 ($\beta_{IBM-1}$) and FRDM ($\beta_{FRDM}$) were the two model dependent values whereas the third one was adopted as the measured value ($\beta_{B(E2)}$) using the $B(E2)\uparrow$ values taken from Refs.~\citep{NNDC22,Pritychenko16}. These measured $\beta_{B(E2)}$ values were calculated using  Eqs. (\ref{def2})-(\ref{BE2up}).

Table~\ref{BGT} compares the total GT strength and centroid values of the selected even-even Fe isotopes calculated using the three different values of deformation parameters (shown in Table~\ref{def}). The Ikeda sum rule was satisfied to around 95$\%$ in our calculation. The current pn-QRPA method considers the residual correlations among nucleons via  (1$p$-1$h$) excitations in large multi-$\hbar\omega$ model spaces. Inclusion of (2$p$-2$h$) or higher excitations may result in 100$\%$ fulfillment of the sum rule. The total GT$_+$ (GT$_-$) values decreases (increases) for heavier isotopes since \textit{ec} (\textit{bd}) becomes challenging (easier). The table shows that the nuclear deformation parameter alters the GT strength and centroid values. These changes later had an impact on the calculated weak interaction rates. It was noted that, in general, higher deformation values resulted in lower GT strength and bigger centroid values for the selected even-even Fe isotope.

As mentioned earlier, we did compute all four weak-interaction mediated rates (\textit{ec}, \textit{bd}, \textit{pc} and \textit{pd}) for each Fe isotope. However, we report only the dominant decay modes (as per our calculations) in this paper. Table~\ref{T1ec} depicts the calculation of pn-QRPA computed \textit{pd} rates (upper panel) for  $^{50 ~\&~ 52}$Fe and \textit{ec} rates (lower panel) for $^{54, ~56 ~\&~ 58}$Fe using the three deformation values at T$_9$ [(1, 3, 5, 10, 20 \& 30) GK] and $\rho$$Y_e$ [($10, 10^{3}$ \& 10$^{5}$) g/cm$^3$]. Table~\ref{T2ec} shows \textit{bd} rates of $^{60-70}$Fe under the same physical conditions. Table~\ref{T3ec} (Table~\ref{T4ec}) presents results similar to that of Table~\ref{T1ec} (Table~\ref{T2ec}) but at higher density values of  $\rho$$Y_e$ [($10^{7}, 10^{9}$ \& 10$^{11}$) g/cm$^3$]. Tables~(\ref{T1ec}-~\ref{T4ec}) show that the computed \textit{pd}, \textit{ec} and \textit{bd} rates on the selected nuclei increase as the core temperature rises. The occupation probability of the parent excited states increases with core temperature, effectively enhancing the overall rates. The \textit{pd} and \textit{ec} rates increase with increasing density of the core material, owing to an increase in electron chemical potential. The \textit{bd} rates, on the other hand, get smaller for high $\rho$$Y_e$ values due to a decrease in available phase space caused by increased electron chemical potential. It is further noted from Tables~(\ref{T1ec}-\ref{T4ec}) that, in general, the weak interaction rates calculated using lower deformation values are bigger up to a few factor than the rates computed with bigger values of nuclear deformation parameter. The reason for this behavior is traced back to the explanation  of calculated GT strength distributions as a function of deformation parameter. We remark caution in establishing a firm correlation between deformation parameters and calculated weak rates. This investigation warrants a much bigger pool of nuclei which we would like to take as a future assignment.

\begin{figure}[h]
	\centering
	\includegraphics[width=1.3\textwidth]{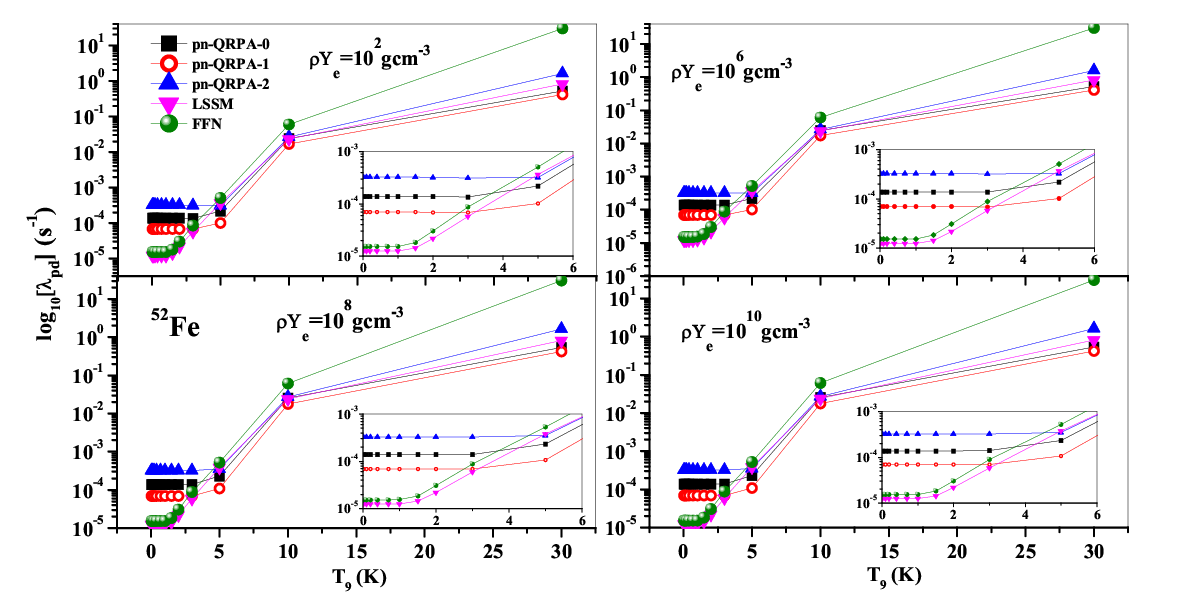}
	\caption{The pn-QRPA computed positron decay rates of $^{52}$Fe in comparison with LSSM and FFN rates for selected core densities and temperatures.  The pn-QRPA-0 refers to the rates computed using deformations from the FRDM, pn-QRPA-1 refers to the rates computed using measured deformations and pn-QRPA-2 refers to the rates computed using deformations from the  IBM-1.} \label{52fe}
\end{figure}

\begin{figure}[h]
	\centering
	\includegraphics[width=1.3\textwidth]{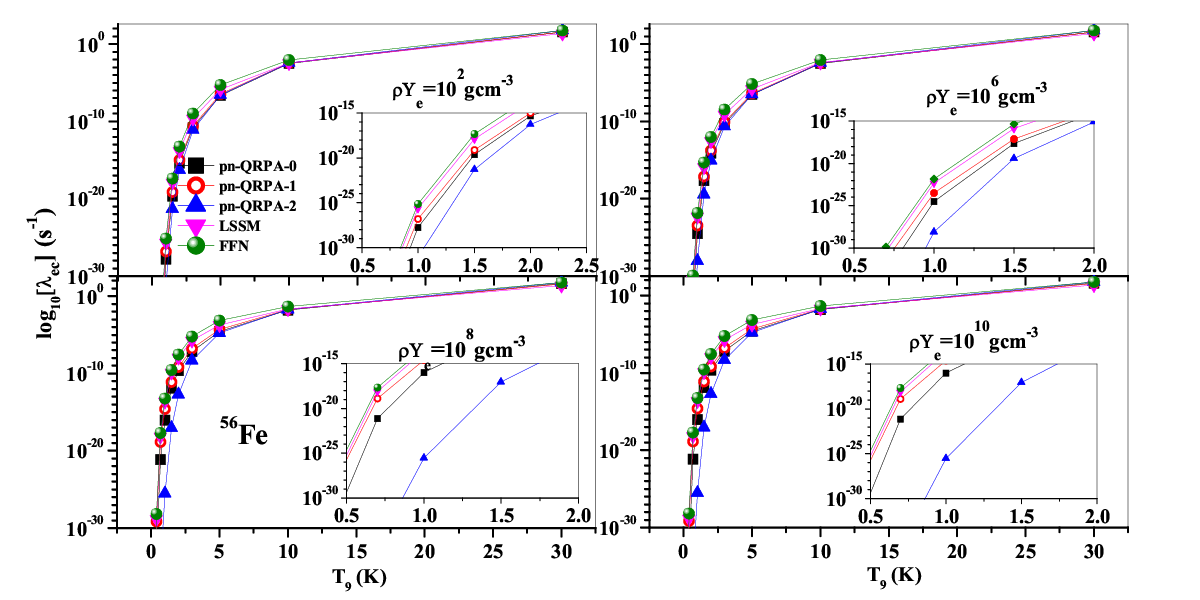}
	\caption{Same as Fig.~\ref{52fe} but for electron capture rates on $^{56}$Fe.} \label{56fe}
\end{figure}

\begin{figure}[h]
	\centering
	\includegraphics[width=1.3\textwidth]{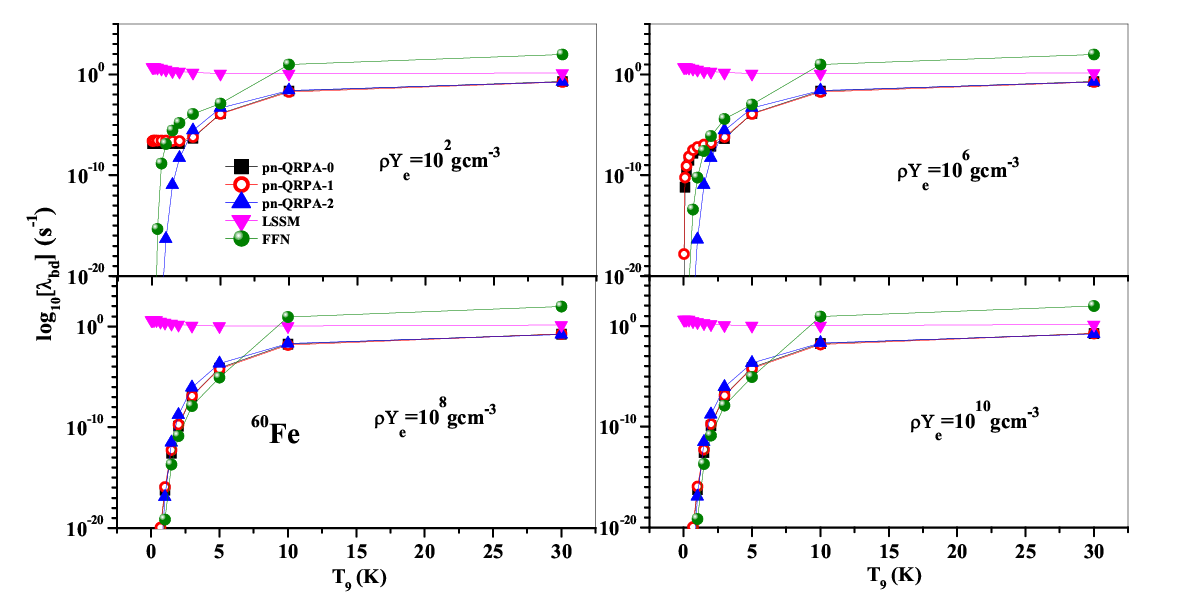}
	\caption{Same as Fig.~\ref{52fe} but for $\beta$-decay rates on $^{60}$Fe.} \label{60fe}
\end{figure}

Fig.~\ref{52fe} depicts the comparison of the pn-QRPA computed positron decay (\textit{pd}) rates for $^{52}$Fe (employing three different values of $\beta$) with the large-scale-shell-model (LSSM) \cite{Lan01} and FFN \cite{Ful82} results.  Figs~(\ref{56fe}-\ref{60fe}) depict a similar comparison of \textit{ec} and \textit{bd} rates for $^{56}$Fe and $^{60}$Fe, respectively. In Figs~(\ref{52fe}-\ref{60fe}), we denote our model calculation using $\beta_{FRDM}$, $\beta_{B(E2)}$ and $\beta_{IBM-1}$ as pn-QRPA-0, pn-QRPA-1 and pn-QRPA-2, respectively. It is noted from Fig.~\ref{52fe} that the pn-QRPA-2 rates are bigger than pn-QRPA-0 and pn-QRPA-1 rates. Spherical deformation led to bigger computed stellar rates. The $\beta_{IBM-1}$ computed higher cumulative GT strengths and lower centroid values which resulted in bigger stellar rates (see Table~\ref{BGT}). In comparison to previous calculations (LSSM and FFN), the pn-QRPA rates are bigger up to an order of magnitude at low temperature range (till 3 GK). Once the core temperature rises, the FFN rates get bigger than pn-QRPA rates up to two orders of magnitude at T$_9$ = 30 GK while LSSM rates are comparable to pn-QRPA positron decay rates of $^{52}$Fe. The pn-QRPA-2 \textit{ec} rates on $^{56}$Fe are smaller than pn-QRPA-0 and pn-QRPA-1 rates (see insets of Fig.~\ref{56fe}) at low core temperatures. Even though Table~\ref{BGT} shows biggest strength and lowest centroid for ground-state GT distribution of $^{56}$Fe computed by the pn-QRPA-2 model, the corresponding \textit{ec} rates are smaller in stellar matter. It is to be noted that the insets show very small magnitudes of rates. The reason for smaller pn-QRPA-2 rates is traced to excited states GT strength distributions which are much different than those shown in Table~\ref{BGT}.    The LSSM agrees well with the three pn-QRPA rates at all densities and temperature values. The FFN computed \textit{ec} rates, however, are bigger up to an order of magnitude than pn-QRPA at high density ($\rho$Y$_e$ = 10 g/cm$^{3}$) and low temperatures (till 3 GK). The reason for this may be traced to misplacement of GT centroid discussed in length in Ref.~\cite{Lan00}. For $^{60}$Fe, the LSSM calculated \textit{bd} rates are up to 3 orders of magnitude bigger than the pn-QRPA results  (Fig.~\ref{60fe}). The difference between pn-QRPA and FFN rates for this case at high T$_9$ values is up to 4 orders of magnitude. The LSSM approach applied the Lanczos method to derive GT strength function but was limited up to 100 iterations which were insufficient for converging the states above 2.5 MeV excitation energies. Moreover, LSSM and FFN calculations used the Brink-Axel hypothesis to estimate rate contributions from high lying parent excited states at high temperatures and densities. These factors led to the differences in pn-QRPA rates with previous calculations which go up to 4 orders of magnitude.
\section{Summary and Conclusion}
We investigated the nuclear structure properties and weak interaction rates of even-even iron isotopes in the mass range A = 50-70. The IBM-1 model was used to compute the energy levels and the B(E2) values of the selected Fe isotopes along the Z = 26 isotopic chain. Their energy ratios were analyzed along the given chain to see behavior of Fe isotopes. The IBM-1 Hamiltonian parameters were fitted by taking these ratios and known $0^+_2$ levels into account. Moreover, their geometric shapes were predicted within the potential energy formalism of the IBM-1 model.  Later, the pn-QRPA model was employed to compute the weak interaction rates of the given Fe isotopes using three different deformation values. It was noted that, in general, bigger deformation values lead to smaller total GT strength and bigger centroid values and subsequently smaller weak rates. The previous calculations are up to 4 orders of magnitude bigger than the reported rates at high core temperatures for even-even isotopes of Fe.

\section*{Acknowledgements}
J.-U. Nabi and M. Riaz would like to acknowledge the support of the Higher Education Commission Pakistan through project
20-15394/NRPU/R\&D/HEC/2021.

\begin{table*}
	\caption{Hamiltonian parameters given in Eq.~(\ref{ham}) in units of MeV. $N$ is the number of bosons and $\sigma$ is defined in Eq. (\ref{rms}). }
	\label{par} \centering
	\begin{tabular}{ccccccccc}
		\hline
		&$N$&$n_{d}$&$a_0$&$a_1$&$a_2$&$\sigma$&$e_b$&$\sigma*$\\ [0.5ex]
		\hline
		$^{50}$Fe&3&1.6322&1.0528&~0.0879&---&0.107&0.102&0\\
		$^{52}$Fe&2&1.6689&0.4653&~0.2193&---&0.087&0.110&1.5\\
		$^{56}$Fe&2&1.2414&---&~0.2051&---&0.013&0.132&1.4\\
		$^{58}$Fe&3&1.0789&---&-0.0092&-0.1232&0.090&0.075&1.8\\
		$^{60}$Fe&4&1.1631&---&-0.0086&-0.0867&0.350&0.054&0.6\\
		$^{62}$Fe&5&1.4632&-0.1537&-0.0994&---&0.852&0.050&0\\
		$^{64}$Fe&6&1.3309&---&-0.0555&---&0.910&---&---\\
		$^{66}$Fe&6&1.1062&---&-0.0505&---&0&---&---\\
		$^{68}$Fe&5&1.0055&---&-0.0683&---&0&---&---\\
		$^{70}$Fe&4&0.7964&---&-0.0827&---&0&---&---\\
		\hline
	\end{tabular}
\end{table*}

\newpage

\begin{table*}[ht]
	\caption{Deformation parameters obtained within the potential energy
		formalism of the IBM-1 and FRDM~\citep{Mol16} models.
		{ Measured $\beta$ values were calculated using the $B(E2)\uparrow$} values taken from Refs.~\citep{NNDC22, Pritychenko16}.}
	\centering
	\begin{tabular}{c c c c }
		\hline\hline
		Nuclei &  Models &$\beta$ & $B(E2)\uparrow$ \\ [0.5ex]
		\hline
		& IBM-1 &-0.189 & --- \\
		$^{50}$Fe & EXP & 0.311 & 0.142 (\emph{70}) \\
		& FRDM & 0.194  & --- \\
		\hline
		& IBM-1 & 0.000 & --- \\
		$^{52}$Fe & EXP & 0.229 & 0.081 (\emph{22}) \\
		& FRDM & 0.118 & --- \\
		\hline
		& IBM-1 & --- & ---  \\		
		$^{54}$Fe &EXP & 0.198  & 0.064 (\emph{3}) \\
		& FRDM & 0.000 & --- \\
		\hline
		& IBM-1 & 0.000 & --- \\
		$^{56}$Fe & EXP & 0.250 & 0.107 (\emph{8}) \\
		& FRDM & 0.117 & --- \\
		\hline
		& IBM-1 & 0.000 & --- \\
		$^{58}$Fe & EXP & 0.262  & 0.123 (\emph{7}) \\
		& FRDM & 0.173 & --- \\
		\hline
		& IBM-1 & 0.000 & --- \\
		$^{60}$Fe & EXP & 0.225 & 0.095 (\emph{16}) \\
		& FRDM & 0.185 & ---  \\
		\hline
		& IBM-1 & 0.000 & ---  \\
		$^{62}$Fe & EXP & 0.229 & 0.103 (\emph{19}) \\
		& FRDM & 0.152  & ---  \\
		\hline
		& IBM-1 & 0.000 & --- \\	
		$^{64}$Fe & EXP & ---  & ---  \\
		& FRDM & -0.084 & ---  \\
		\hline
		& IBM-1 & 0.000 \\	
		$^{66}$Fe & EXP & --- & --- \\
		& FRDM & 0.000 & ---  \\
		\hline
		& IBM-1 & 0.469 & --- \\	
		$^{68}$Fe & EXP & --- & ---  \\
		& FRDM & 0.000 & ---  \\
		\hline
		& IBM-1 & 0.555 & --- \\	
		$^{70}$Fe & EXP & --- & --- \\
		& FRDM & 0.128 & --- \\
		\hline
	\end{tabular}
	\label{def}
\end{table*}

\newpage
\begin{table*}[]
	\centering\scriptsize\caption{Statistical data of the pn-QRPA computed GT strength distributions using the three deformation values ($\beta_{FRDM}$, $\beta_{B(E2)}$ \& $\beta_{IBM-1}$) for $^{50-70}$Fe.} \label{BGT}
	\renewcommand{\arraystretch}{1.2}
	\begin{tabular}{|c|ccc|ccc|}
		\hline
		\hline
		\multicolumn{1}{l}{}       & \multicolumn{3}{c}{$\sum$ GT$_+$}                                                      & \multicolumn{3}{c}{Centroid GT$_+$~~(MeV)}                                                          \\
		\hline
		\multicolumn{1}{l}{Nuclei} & \multicolumn{1}{l}{$\sum$ GT$_{\beta_{FRDM}}$} & \multicolumn{1}{l}{$\sum$ GT$_{\beta_{B(E2)}}$} & \multicolumn{1}{l}{$\sum$ GT$_{\beta_{IBM-1}}$} & \multicolumn{1}{l}{$\bar{E}_{\beta_{FRDM}}$} & \multicolumn{1}{l}{$\bar{E}_{\beta_{B(E2)}}$} & \multicolumn{1}{l}{$\bar{E}_{\beta_{IBM-1}}$}  \\
		\hline
		$^{50}$Fe & 12.57 & 12.22 & 12.80 & 8.59 & 8.56 & 8.50  \\
		$^{52}$Fe & 10.55 & 10.08 & 10.76 & 7.34 & 7.33 & 7.27  \\
		$^{54}$Fe & 8.58  & 7.99  &  ---  & 5.26 & 5.95 &  ---  \\
		$^{56}$Fe & 8.21  & 7.27  & 8.47  & 4.32 & 4.48 & 4.26  \\
		$^{58}$Fe & 7.84  & 7.25  & 8.72  & 2.63 & 2.81 & 2.40  \\
		$^{60}$Fe & 2.69  & 2.50  & 3.49  & 3.49 & 3.88 & 2.40  \\
		$^{62}$Fe & 1.87  & 1.78  & 2.18  & 3.08 & 3.78 & 2.58  \\
		$^{64}$Fe & 1.15  &  ---  & 1.20  & 3.03 &  --- & 3.19  \\
		$^{66}$Fe & 0.64  &  ---  & 0.64  & 4.89 &  --- & 4.89  \\
		$^{68}$Fe & 0.37  &  ---  & 0.51  & 8.57 &  --- & 6.13  \\
		$^{70}$Fe & 0.31  &  ---  & 0.47  & 8.35 &  --- & 6.18  \\
		\hline
		\multicolumn{1}{l}{}       & \multicolumn{3}{c}{$\sum$ GT$_-$}                                                      & \multicolumn{3}{c}{Centroid GT$_-$~~(MeV)}                                                            \\
		\hline
		$^{50}$Fe & 6.28  & 5.95  & 6.52  & 4.68  & 4.58  & 5.09    \\
		$^{52}$Fe & 10.28 & 9.81  & 10.48 & 6.10  & 6.02  & 6.27    \\
		$^{54}$Fe & 14.30 & 13.70 &  ---  & 7.26  & 7.15  & ---     \\
		$^{56}$Fe & 19.66 & 18.72 & 19.92 & 8.53  & 8.02  & 8.93     \\
		$^{58}$Fe & 25.01 & 24.42 & 25.91 & 9.37  & 9.15  & 9.46     \\
		$^{60}$Fe & 26.67 & 26.48 & 27.49 & 11.45 & 11.38 & 11.98    \\
		$^{62}$Fe & 31.84 & 31.70 & 31.05 & 12.82 & 12.60 & 12.81    \\
		$^{64}$Fe & 37.09 &  ---  & 36.33 & 13.96 &  ---  & 12.09    \\
		$^{66}$Fe & 42.64 &  ---  & 42.64 & 14.52 &  ---  & 14.52    \\
		$^{68}$Fe & 48.36 &  ---  & 48.41 & 15.93 &  ---  & 16.31    \\
		$^{70}$Fe & 54.26 &  ---  & 50.94 & 17.13 &  ---  & 17.27    \\
		\hline
	\end{tabular}
\end{table*}

\begin{table*}[]
	\scriptsize\caption{Positron decay  (\textit{pd}) rates of $^{50-52}$Fe (upper panel) and \textit{ec} rates on $^{54-58}$Fe (lower panel) calculated using the three deformation values ($\beta_{FRDM}$, $\beta_{B(E2)}$ \& $\beta_{IBM-1}$). The units of $T_9$, $\rho$$\it Y_{e}$ and $\lambda^{ec}$/$\lambda^{pd}$ are GK, g/cm$^3$ and $s^{-1}$, respectively. The exponents are shown in parenthesis.}
	\label{T1ec}
	\centering
	\scalebox{.90}{
		\begin{tabular}{c|c|ccc|ccc|ccc|}
			\hline
			\multicolumn{5}{l}{} & \multicolumn{1}{l}{$\lambda^{pd}$} & \multicolumn{5}{l}{}        \\
			\cline{1-11} \multicolumn{1}{l}{} &  & \multicolumn{3}{c}{$\rho$$\it Y_{e}$ = $10^1$ }        & \multicolumn{3}{c}{$\rho$$\it Y_{e}$ = $10^3$ }& \multicolumn{3}{c}{$\rho$$\it Y_{e}$ = $10^{5}$ }         \\
			\cline{3-11} \multicolumn{1}{l}{Nuclei} &       $T_9$ & $\lambda_{\beta_{FRDM}}$& $\lambda_{\beta_{B(E2)}}$    & $\lambda_{\beta_{IBM-1}}$      &$\lambda_{\beta_{FRDM}}$ & $\lambda_{\beta_{B(E2)}}$    & $\lambda_{\beta_{IBM-1}}$      & $\lambda_{\beta_{FRDM}}$& $\lambda_{\beta_{B(E2)}}$    & $\lambda_{\beta_{IBM-1}}$ \\
			\hline\\
			$^{50}$Fe   & 1   & 1.13(+00) & 9.77(-01) & 1.51(+00) & 1.13(+00) & 9.77(-01) & 1.51(+00) & 1.13(+00) & 9.77(-01) & 1.51(+00) \\
			& 3   & 1.13(+00) & 9.75(-01) & 1.51(+00) & 1.13(+00) & 9.75(-01) & 1.51(+00) & 1.13(+00) & 9.75(-01) & 1.51(+00) \\
			& 5   & 1.13(+00) & 9.73(-01) & 1.51(+00) & 1.13(+00) & 9.73(-01) & 1.51(+00) & 1.13(+00) & 9.73(-01) & 1.51(+00) \\
			& 10  & 1.81(+00) & 1.58(+00) & 2.38(+00) & 1.81(+00) & 1.58(+00) & 2.38(+00) & 1.81(+00) & 1.58(+00) & 2.38(+00) \\
			& 20  & 7.52(+00) & 8.30(+00) & 7.52(+00) & 7.52(+00) & 8.30(+00) & 7.52(+00) & 7.52(+00) & 8.30(+00) & 7.52(+00) \\
			& 30  & 1.00(+01) & 1.18(+01) & 9.93(+00) & 1.00(+01) & 1.18(+01) & 9.93(+00) & 1.00(+01) & 1.18(+01) & 9.93(+00) \\
			&&&&&&&&&&\\
			$^{52}$Fe   & 1   & 1.39(-04) & 6.92(-05) & 3.24(-04) & 1.39(-04) & 6.92(-05) & 3.24(-04) & 1.39(-04) & 6.92(-05) & 3.24(-04) \\
			& 3   & 1.35(-04) & 6.73(-05) & 3.14(-04) & 1.35(-04) & 6.73(-05) & 3.14(-04) & 1.35(-04) & 6.76(-05) & 3.16(-04) \\
			& 5   & 2.17(-04) & 1.00(-04) & 3.17(-04) & 2.17(-04) & 1.00(-04) & 3.17(-04) & 2.17(-04) & 1.01(-04) & 3.17(-04) \\
			& 10  & 2.43(-02) & 1.70(-02) & 2.62(-02) & 2.43(-02) & 1.70(-02) & 2.62(-02) & 2.43(-02) & 1.70(-02) & 2.62(-02) \\
			& 20  & 3.24(-01) & 2.77(-01) & 7.13(-01) & 3.24(-01) & 2.77(-01) & 7.13(-01) & 3.24(-01) & 2.77(-01) & 7.13(-01) \\
			& 30  & 5.25(-01) & 4.14(-01) & 1.60(+00) & 5.25(-01) & 4.14(-01) & 1.60(+00) & 5.25(-01) & 4.14(-01) & 1.60(+00) \\
			&&&&&&&&&&\\
			\hline
			\multicolumn{5}{l}{} & \multicolumn{1}{l}{$\lambda^{ec}$} & \multicolumn{5}{l}{}        \\
			\hline
			&&&&&&&&&&\\
			$^{54}$Fe   & 1   & 1.55(-13) & 5.22(-13) &   ---     & 2.61(-13) & 8.83(-13) &   ---     & 1.87(-11) & 6.31(-11) &   ---     \\
			& 3   & 6.00(-07) & 4.72(-07) &   ---     & 6.01(-07) & 4.73(-07) &   ---     & 6.89(-07) & 5.42(-07) &   ---     \\
			& 5   & 4.60(-05) & 4.58(-05) &   ---     & 4.60(-05) & 4.58(-05) &   ---     & 4.70(-05) & 4.68(-05) &   ---     \\
			& 10  & 3.40(-02) & 2.26(-02) &   ---     & 3.41(-02) & 2.26(-02) &   ---     & 3.42(-02) & 2.26(-02) &   ---     \\
			& 20  & 1.19(+01) & 4.57(+00) &   ---     & 1.19(+01) & 4.58(+00) &   ---     & 1.19(+01) & 4.58(+00) &   ---     \\
			& 30  & 1.68(+02) & 5.94(+01) &   ---     & 1.68(+02) & 5.96(+01) &   ---     & 1.68(+02) & 5.96(+01) &   ---     \\
			&&&&&&&&&&\\
			$^{56}$Fe   & 1   & 1.52(-28) & 1.43(-27) & 8.07(-32) & 2.58(-28) & 2.42(-27) & 1.36(-31) & 1.79(-26) & 1.72(-25) & 8.77(-30) \\
			& 3   & 1.60(-11) & 2.72(-11) & 7.83(-12) & 1.60(-11) & 2.73(-11) & 7.85(-12) & 1.83(-11) & 3.12(-11) & 8.95(-12) \\
			& 5   & 2.39(-07) & 3.54(-07) & 2.54(-07) & 2.39(-07) & 3.55(-07) & 2.54(-07) & 2.44(-07) & 3.62(-07) & 2.59(-07) \\
			& 10  & 3.03(-03) & 3.35(-03) & 3.74(-03) & 3.03(-03) & 3.36(-03) & 3.74(-03) & 3.04(-03) & 3.37(-03) & 3.75(-03) \\
			& 20  & 2.03(+00) & 1.90(+00) & 2.92(+00) & 2.03(+00) & 1.90(+00) & 2.92(+00) & 2.03(+00) & 1.90(+00) & 2.92(+00) \\
			& 30  & 3.97(+01) & 3.65(+01) & 5.66(+01) & 3.97(+01) & 3.65(+01) & 5.66(+01) & 3.97(+01) & 3.65(+01) & 5.66(+01) \\
			&&&&&&&&&&\\
			$^{58}$Fe   & 1   & 1.94(-39) & 1.31(-39) & 1.87(-39) & 3.27(-39) & 2.21(-39) & 3.17(-39) & 2.33(-37) & 1.58(-37) & 2.26(-37) \\
			& 3   & 1.25(-14) & 1.30(-14) & 9.98(-15) & 1.25(-14) & 1.31(-14) & 9.98(-15) & 1.44(-14) & 1.50(-14) & 1.14(-14) \\
			& 5   & 7.38(-09) & 7.71(-09) & 6.38(-09) & 7.38(-09) & 7.71(-09) & 6.38(-09) & 7.55(-09) & 7.87(-09) & 6.52(-09) \\
			& 10  & 8.49(-04) & 8.26(-04) & 7.85(-04) & 8.49(-04) & 8.28(-04) & 7.85(-04) & 8.51(-04) & 8.30(-04) & 7.87(-04) \\
			& 20  & 1.18(+00) & 1.14(+00) & 1.23(+00) & 1.19(+00) & 1.14(+00) & 1.23(+00) & 1.19(+00) & 1.14(+00) & 1.23(+00) \\
			& 30  & 2.78(+01) & 2.69(+01) & 2.59(+01) & 2.78(+01) & 2.69(+01) & 2.59(+01) & 2.78(+01) & 2.69(+01) & 2.59(+01) \\
			&&&&&&&&&&\\			
			\hline
	\end{tabular}}
\end{table*}
 
\begin{table*}[]
	\scriptsize\caption{ $\beta$-decay (\textit{bd}) rates of $^{60-70}$Fe calculated using the three deformation values ($\beta_{FRDM}$, $\beta_{B(E2)}$ \& $\beta_{IBM-1}$). The units of $T_9$, $\rho$$\it Y_{e}$ and $\lambda^{bd}$ are GK, g/cm$^3$ and $s^{-1}$, respectively. The exponents are shown in parenthesis.}
	\label{T2ec}
	\centering
	\scalebox{0.90}{
		\begin{tabular}{c|c|ccc|ccc|ccc|}
			\hline
			\multicolumn{5}{l}{} & \multicolumn{1}{l}{$\lambda^{bd}$} & \multicolumn{5}{l}{}        \\
			\cline{1-11} \multicolumn{1}{l}{} &  & \multicolumn{3}{c}{$\rho$$\it Y_{e}$ = $10^1$ }        & \multicolumn{3}{c}{$\rho$$\it Y_{e}$ = $10^3$ }& \multicolumn{3}{c}{$\rho$$\it Y_{e}$ = $10^{5}$ }         \\
			\cline{3-11} \multicolumn{1}{l}{Nuclei} &       $T_9$ & $\lambda_{\beta_{FRDM}}$& $\lambda_{\beta_{B(E2)}}$    & $\lambda_{\beta_{IBM-1}}$      &$\lambda_{\beta_{FRDM}}$ & $\lambda_{\beta_{B(E2)}}$    & $\lambda_{\beta_{IBM-1}}$      & $\lambda_{\beta_{FRDM}}$& $\lambda_{\beta_{B(E2)}}$    & $\lambda_{\beta_{IBM-1}}$ \\
			\hline\\
			$^{60}$Fe   & 1   & 1.40(-07) & 2.62(-07) & 4.51(-17) & 1.40(-07) & 2.61(-07) & 4.51(-17) & 1.18(-07) & 2.22(-07) & 4.49(-17) \\
			& 3   & 4.86(-07) & 6.31(-07) & 2.64(-06) & 4.86(-07) & 6.31(-07) & 2.64(-06) & 4.84(-07) & 6.27(-07) & 2.64(-06) \\
			& 5   & 1.39(-04) & 1.15(-04) & 4.53(-04) & 1.39(-04) & 1.15(-04) & 4.53(-04) & 1.39(-04) & 1.15(-04) & 4.53(-04) \\
			& 10  & 2.25(-02) & 1.87(-02) & 2.61(-02) & 2.25(-02) & 1.87(-02) & 2.61(-02) & 2.25(-02) & 1.87(-02) & 2.61(-02) \\
			& 20  & 1.44(-01) & 1.27(-01) & 1.39(-01) & 1.44(-01) & 1.27(-01) & 1.39(-01) & 1.44(-01) & 1.27(-01) & 1.39(-01) \\
			& 30  & 1.87(-01) & 1.71(-01) & 1.67(-01) & 1.87(-01) & 1.71(-01) & 1.67(-01) & 1.87(-01) & 1.71(-01) & 1.67(-01) \\
			&&&&&&&&&&\\
			$^{62}$Fe   & 1   & 3.40(-03) & 3.05(-03) & 7.57(-03) & 3.40(-03) & 3.05(-03) & 7.55(-03) & 3.36(-03) & 3.02(-03) & 7.50(-03) \\
			& 3   & 3.34(-03) & 3.00(-03) & 7.50(-03) & 3.34(-03) & 3.00(-03) & 7.50(-03) & 3.33(-03) & 2.99(-03) & 7.48(-03) \\
			& 5   & 5.75(-03) & 4.20(-03) & 1.30(-02) & 5.75(-03) & 4.20(-03) & 1.30(-02) & 5.74(-03) & 4.20(-03) & 1.30(-02) \\
			& 10  & 2.37(-01) & 1.53(-01) & 2.77(-01) & 2.37(-01) & 1.53(-01) & 2.77(-01) & 2.37(-01) & 1.53(-01) & 2.77(-01) \\
			& 20  & 1.15(+00) & 8.73(-01) & 1.71(+00) & 1.15(+00) & 8.73(-01) & 1.71(+00) & 1.15(+00) & 8.73(-01) & 1.71(+00) \\
			& 30  & 1.50(+00) & 1.10(+00) & 2.26(+00) & 1.50(+00) & 1.10(+00) & 2.26(+00) & 1.50(+00) & 1.10(+00) & 2.26(+00) \\
			&&&&&&&&&&\\
			$^{64}$Fe   & 1   & 3.53(-01) &   ---     & 4.58(-01) & 3.53(-01) &   ---     & 4.58(-01) & 3.52(-01) &   ---     & 4.58(-01) \\
			& 3   & 3.52(-01) &   ---     & 4.57(-01) & 3.52(-01) &   ---     & 4.57(-01) & 3.52(-01) &   ---     & 4.57(-01) \\
			& 5   & 3.74(-01) &   ---     & 4.82(-01) & 3.74(-01) &   ---     & 4.82(-01) & 3.73(-01) &   ---     & 4.81(-01) \\
			& 10  & 1.33(+00) &   ---     & 1.78(+00) & 1.33(+00) &   ---     & 1.78(+00) & 1.33(+00) &   ---     & 1.78(+00) \\
			& 20  & 2.61(+00) &   ---     & 7.06(+00) & 2.61(+00) &   ---     & 7.06(+00) & 2.61(+00) &   ---     & 7.06(+00) \\
			& 30  & 2.64(+00) &   ---     & 8.13(+00) & 2.64(+00) &   ---     & 8.13(+00) & 2.64(+00) &   ---     & 8.13(+00) \\
			&&&&&&&&&&\\
			$^{66}$Fe   & 1   & 2.62(+00) &   ---     & 2.62(+00) & 2.62(+00) &   ---     & 2.62(+00) & 2.61(+00) &   ---     & 2.61(+00) \\
			& 3   & 2.61(+00) &   ---     & 2.61(+00) & 2.61(+00) &   ---     & 2.61(+00) & 2.61(+00) &   ---     & 2.61(+00) \\
			& 5   & 2.61(+00) &   ---     & 2.61(+00) & 2.61(+00) &   ---     & 2.61(+00) & 2.61(+00) &   ---     & 2.61(+00) \\
			& 10  & 4.58(+00) &   ---     & 4.58(+00) & 4.58(+00) &   ---     & 4.58(+00) & 4.58(+00) &   ---     & 4.58(+00) \\
			& 20  & 2.42(+01) &   ---     & 2.42(+01) & 2.42(+01) &   ---     & 2.42(+01) & 2.42(+01) &   ---     & 2.42(+01) \\
			& 30  & 3.47(+01) &   ---     & 3.47(+01) & 3.47(+01) &   ---     & 3.47(+01) & 3.47(+01) &   ---     & 3.47(+01) \\
			&&&&&&&&&&\\
			$^{68}$Fe   & 1   & 5.22(+00) &   ---     & 3.32(+00) & 5.22(+00) &   ---     & 3.32(+00) & 5.22(+00) &   ---     & 3.31(+00) \\
			& 3   & 5.22(+00) &   ---     & 3.31(+00) & 5.22(+00) &   ---     & 3.31(+00) & 5.22(+00) &   ---     & 3.31(+00) \\
			& 5   & 5.21(+00) &   ---     & 3.31(+00) & 5.21(+00) &   ---     & 3.31(+00) & 5.21(+00) &   ---     & 3.31(+00) \\
			& 10  & 8.65(+00) &   ---     & 4.73(+00) & 8.65(+00) &   ---     & 4.73(+00) & 8.65(+00) &   ---     & 4.73(+00) \\
			& 20  & 5.05(+01) &   ---     & 9.48(+00) & 5.05(+01) &   ---     & 9.48(+00) & 5.05(+01) &   ---     & 9.48(+00) \\
			& 30  & 7.46(+01) &   ---     & 1.07(+01) & 7.46(+01) &   ---     & 1.07(+01) & 7.46(+01) &   ---     & 1.07(+01) \\
			&&&&&&&&&&\\
			$^{70}$Fe   & 1   & 1.47(+01) &   ---     & 1.02(+01) & 1.47(+01) &   ---     & 1.02(+01) & 1.47(+01) &   ---     & 1.02(+01) \\
			& 3   & 1.47(+01) &   ---     & 1.02(+01) & 1.47(+01) &   ---     & 1.02(+01) & 1.47(+01) &   ---     & 1.02(+01) \\
			& 5   & 1.48(+01) &   ---     & 1.03(+01) & 1.48(+01) &   ---     & 1.03(+01) & 1.48(+01) &   ---     & 1.03(+01) \\
			& 10  & 1.92(+01) &   ---     & 1.40(+01) & 1.92(+01) &   ---     & 1.40(+01) & 1.92(+01) &   ---     & 1.40(+01) \\
			& 20  & 2.82(+01) &   ---     & 2.63(+01) & 2.82(+01) &   ---     & 2.63(+01) & 2.82(+01) &   ---     & 2.63(+01) \\
			& 30  & 3.08(+01) &   ---     & 2.90(+01) & 3.08(+01) &   ---     & 2.90(+01) & 3.08(+01) &   ---     & 2.90(+01) \\
			&&&&&&&&&&\\
			\\ \hline
	\end{tabular}}
\end{table*}

\begin{table*}[]
	\scriptsize\caption{Same as Table~\ref{T1ec} but at densities $\rho$$\it Y_{e}$ = $10^7$, $10^9$ \& $10^{11}$.}
	\label{T3ec}
	\centering
	\scalebox{0.90}{
		\begin{tabular}{c|c|ccc|ccc|ccc|}
			\hline
			\multicolumn{5}{l}{} & \multicolumn{1}{l}{$\lambda^{pd}$} & \multicolumn{5}{l}{}        \\
			\cline{1-11} \multicolumn{1}{l}{} &  & \multicolumn{3}{c}{$\rho$$\it Y_{e}$ = $10^7$ }        & \multicolumn{3}{c}{$\rho$$\it Y_{e}$ = $10^9$ }& \multicolumn{3}{c}{$\rho$$\it Y_{e}$ = $10^{11}$ }         \\
			\cline{3-11} \multicolumn{1}{l}{Nuclei} &       $T_9$ & $\lambda_{\beta_{FRDM}}$& $\lambda_{\beta_{B(E2)}}$    & $\lambda_{\beta_{IBM-1}}$      &$\lambda_{\beta_{FRDM}}$ & $\lambda_{\beta_{B(E2)}}$    & $\lambda_{\beta_{IBM-1}}$      & $\lambda_{\beta_{FRDM}}$& $\lambda_{\beta_{B(E2)}}$    & $\lambda_{\beta_{IBM-1}}$ \\
			\hline\\
			$^{50}$Fe   & 1   & 1.13(+00) & 9.77(-01) & 1.51(+00) & 1.13(+00) & 9.77(-01) & 1.51(+00) & 1.13(+00)  & 9.77(-01)  & 1.51(+00) \\
			& 3   & 1.13(+00) & 9.77(-01) & 1.51(+00) & 1.13(+00) & 9.77(-01) & 1.51(+00) & 1.13(+00)  & 9.77(-01)  & 1.51(+00) \\
			& 5   & 1.14(+00) & 9.79(-01) & 1.52(+00) & 1.14(+00) & 9.82(-01) & 1.52(+00) & 1.14(+00)  & 9.82(-01)  & 1.52(+00) \\
			& 10  & 1.82(+00) & 1.60(+00) & 2.39(+00) & 1.87(+00) & 1.64(+00) & 2.45(+00) & 1.87(+00)  & 1.64(+00)  & 2.45(+00) \\
			& 20  & 7.53(+00) & 8.30(+00) & 7.52(+00) & 8.04(+00) & 8.83(+00) & 8.04(+00) & 8.13(+00)  & 8.95(+00)  & 8.15(+00) \\
			& 30  & 1.00(+01) & 1.18(+01) & 9.95(+00) & 1.07(+01) & 1.26(+01) & 1.07(+01) & 1.15(+01)  & 1.35(+01)  & 1.15(+01) \\
			&&&&&&&&&&\\
			$^{52}$Fe   & 1   & 1.39(-04) & 6.92(-05) & 3.24(-04) & 1.39(-04) & 6.92(-05) & 3.24(-04) & 1.39(-04)  & 6.92(-05)  & 3.24(-04) \\
			& 3   & 1.39(-04) & 6.92(-05) & 3.24(-04) & 1.39(-04) & 6.92(-05) & 3.24(-04) & 1.39(-04)  & 6.92(-05)  & 3.24(-04) \\
			& 5   & 2.29(-04) & 1.06(-04) & 3.43(-04) & 2.32(-04) & 1.07(-04) & 3.50(-04) & 2.32(-04)  & 1.07(-04)  & 3.50(-04) \\
			& 10  & 2.45(-02) & 1.72(-02) & 2.64(-02) & 2.56(-02) & 1.79(-02) & 2.72(-02) & 2.56(-02)  & 1.79(-02)  & 2.72(-02) \\
			& 20  & 3.25(-01) & 2.78(-01) & 7.16(-01) & 3.66(-01) & 3.16(-01) & 8.02(-01) & 3.75(-01)  & 3.24(-01)  & 8.18(-01) \\
			& 30  & 5.26(-01) & 4.15(-01) & 1.60(+00) & 5.86(-01) & 4.67(-01) & 1.79(+00) & 6.56(-01)  & 5.28(-01)  & 2.01(+00) \\
			&&&&&&&&&&\\
			\hline
			\multicolumn{5}{l}{} & \multicolumn{1}{l}{$\lambda^{ec}$} & \multicolumn{5}{l}{}        \\
			\hline
			&&&&&&&&&&\\
			$^{54}$Fe   & 1   & 1.72(-07) & 5.45(-07) &   ---     & 3.11(-01) & 5.09(-01) &   ---     & 4.81(+04)  & 4.05(+04)  &   ---      \\
			& 3   & 3.08(-05) & 2.39(-05) &   ---     & 3.56(-01) & 5.82(-01) &   ---     & 4.82(+04)  & 4.06(+04)  &   ---      \\
			& 5   & 2.29(-04) & 2.28(-04) &   ---     & 5.50(-01) & 7.73(-01) &   ---     & 4.84(+04)  & 4.07(+04)  &   ---      \\
			& 10  & 4.25(-02) & 2.82(-02) &   ---     & 4.04(+00) & 3.06(+00) &   ---     & 5.13(+04)  & 4.23(+04)  &   ---      \\
			& 20  & 1.22(+01) & 4.70(+00) &   ---     & 7.52(+01) & 3.03(+01) &   ---     & 8.95(+04)  & 5.25(+04)  &   ---      \\
			& 30  & 1.69(+02) & 6.00(+01) &   ---     & 3.51(+02) & 1.26(+02) &   ---     & 1.36(+05)  & 6.37(+04)  &   ---      \\
			&&&&&&&&&&\\
			$^{56}$Fe   & 1   & 9.73(-23) & 1.46(-21) & 9.53(-28) & 5.50(-04) & 9.57(-04) & 6.68(-13) & 3.27(+04)  & 2.84(+04)  & 3.43(+04) \\
			& 3   & 6.93(-10) & 1.30(-09) & 2.38(-10) & 1.81(-03) & 4.25(-03) & 6.30(-06) & 3.27(+04)  & 2.85(+04)  & 3.44(+04) \\
			& 5   & 1.15(-06) & 1.75(-06) & 1.16(-06) & 8.77(-03) & 1.89(-02) & 1.94(-03) & 3.30(+04)  & 2.86(+04)  & 3.46(+04) \\
			& 10  & 3.80(-03) & 4.20(-03) & 4.68(-03) & 4.73(-01) & 5.48(-01) & 5.09(-01) & 3.44(+04)  & 2.99(+04)  & 3.68(+04) \\
			& 20  & 2.08(+00) & 1.95(+00) & 3.01(+00) & 1.39(+01) & 1.30(+01) & 1.99(+01) & 4.56(+04)  & 3.98(+04)  & 5.82(+04) \\
			& 30  & 4.00(+01) & 3.68(+01) & 5.71(+01) & 8.45(+01) & 7.76(+01) & 1.20(+02) & 5.79(+04)  & 5.20(+04)  & 7.78(+04) \\
			&&&&&&&&&&\\
			$^{58}$Fe   & 1   & 2.07(-33) & 1.40(-33) & 2.09(-33) & 6.85(-14) & 1.31(-14) & 4.90(-18) & 2.70(+04)  & 2.44(+04)  & 3.16(+04) \\
			& 3   & 6.40(-13) & 6.70(-13) & 5.19(-13) & 2.25(-06) & 2.75(-06) & 1.07(-06) & 2.72(+04)  & 2.45(+04)  & 3.16(+04) \\
			& 5   & 3.72(-08) & 3.88(-08) & 3.23(-08) & 6.53(-04) & 7.10(-04) & 5.74(-04) & 2.73(+04)  & 2.47(+04)  & 3.18(+04) \\
			& 10  & 1.06(-03) & 1.04(-03) & 9.86(-04) & 1.79(-01) & 1.75(-01) & 1.76(-01) & 2.84(+04)  & 2.59(+04)  & 3.34(+04) \\
			& 20  & 1.22(+00) & 1.17(+00) & 1.26(+00) & 8.34(+00) & 8.02(+00) & 8.69(+00) & 3.64(+04)  & 3.46(+04)  & 4.03(+04) \\
			& 30  & 2.80(+01) & 2.70(+01) & 2.62(+01) & 5.93(+01) & 5.73(+01) & 5.55(+01) & 4.61(+04)  & 4.44(+04)  & 4.35(+04) \\
			&&&&&&&&&&\\		
			\hline
	\end{tabular}}
\end{table*}

\begin{table*}[]
	\scriptsize\caption{Same as Table~\ref{T2ec} but at densities $\rho$$\it Y_{e}$ = $10^7$, $10^9$ \& $10^{11}$. }
	\label{T4ec}
	\centering
	\scalebox{0.90}{
		\begin{tabular}{c|c|ccc|ccc|ccc|}
			\hline
			\multicolumn{5}{l}{} & \multicolumn{1}{l}{$\lambda^{bd}$} & \multicolumn{5}{l}{}        \\
			\cline{1-11} \multicolumn{1}{l}{} &  & \multicolumn{3}{c}{$\rho$$\it Y_{e}$ = $10^7$ }        & \multicolumn{3}{c}{$\rho$$\it Y_{e}$ = $10^9$ }& \multicolumn{3}{c}{$\rho$$\it Y_{e}$ = $10^{11}$ }         \\
			\cline{3-11} \multicolumn{1}{l}{Nuclei} &       $T_9$ & $\lambda_{\beta_{FRDM}}$& $\lambda_{\beta_{B(E2)}}$    & $\lambda_{\beta_{IBM-1}}$      &$\lambda_{\beta_{FRDM}}$ & $\lambda_{\beta_{B(E2)}}$    & $\lambda_{\beta_{IBM-1}}$      & $\lambda_{\beta_{FRDM}}$& $\lambda_{\beta_{B(E2)}}$    & $\lambda_{\beta_{IBM-1}}$ \\
			\hline\\
			$^{60}$Fe   & 1   & 9.77(-11) & 1.97(-10) & 3.78(-17) & 6.78(-27) & 2.68(-27) & 1.62(-26) & 1.00(-100) & 1.00(-100) & 1.00(-100) \\
			& 3   & 3.34(-07) & 3.78(-07) & 2.30(-06) & 3.72(-10) & 2.05(-10) & 5.15(-10) & 1.75(-41)  & 1.06(-41)  & 1.46(-41) \\
			& 5   & 1.30(-04) & 1.07(-04) & 4.18(-04) & 2.19(-06) & 1.56(-06) & 2.21(-06) & 3.56(-25)  & 2.60(-25)  & 2.37(-25) \\
			& 10  & 2.22(-02) & 1.83(-02) & 2.55(-02) & 3.00(-03) & 2.55(-03) & 2.09(-03) & 1.17(-12)  & 9.77(-13)  & 5.70(-13) \\
			& 20  & 1.43(-01) & 1.26(-01) & 1.38(-01) & 7.41(-02) & 6.58(-02) & 6.19(-02) & 1.41(-06)  & 1.22(-06)  & 8.73(-07) \\
			& 30  & 1.87(-01) & 1.70(-01) & 1.67(-01) & 1.47(-01) & 1.34(-01) & 1.25(-01) & 1.22(-04)  & 1.09(-04)  & 8.17(-05) \\
			&&&&&&&&&&\\
			$^{62}$Fe   & 1   & 1.92(-03) & 1.75(-03) & 4.49(-03) & 1.67(-17) & 3.57(-18) & 2.82(-16) & 1.00(-100) & 1.00(-100) & 1.00(-100) \\
			& 3   & 2.29(-03) & 2.06(-03) & 5.25(-03) & 4.94(-07) & 2.42(-07) & 2.00(-06) & 7.08(-37)  & 8.93(-38)  & 4.27(-37) \\
			& 5   & 5.11(-03) & 3.64(-03) & 1.16(-02) & 2.79(-04) & 1.34(-04) & 4.28(-04) & 3.78(-22)  & 8.02(-23)  & 1.91(-22) \\
			& 10  & 2.34(-01) & 1.52(-01) & 2.74(-01) & 6.28(-02) & 3.98(-02) & 6.34(-02) & 8.32(-11)  & 2.90(-11)  & 3.76(-11) \\
			& 20  & 1.15(+00) & 8.69(-01) & 1.70(+00) & 7.21(-01) & 5.25(-01) & 9.79(-01) & 2.78(-05)  & 1.40(-05)  & 2.23(-05) \\
			& 30  & 1.50(+00) & 1.10(+00) & 2.26(+00) & 1.25(+00) & 9.04(-01) & 1.82(+00) & 1.67(-03)  & 9.42(-04)  & 1.68(-03) \\
			&&&&&&&&&&\\
			$^{64}$Fe   & 1   & 3.17(-01) &   ---     & 4.15(-01) & 1.50(-05) &   ---     & 2.30(-05) & 1.05(-97)  &    ---     & 5.81(-97) \\
			& 3   & 3.22(-01) &   ---     & 4.22(-01) & 8.65(-04) &   ---     & 1.66(-03) & 1.87(-33)  &    ---     & 4.62(-33) \\
			& 5   & 3.54(-01) &   ---     & 4.58(-01) & 1.30(-02) &   ---     & 1.67(-02) & 3.66(-20)  &    ---     & 6.43(-20) \\
			& 10  & 1.32(+00) &   ---     & 1.76(+00) & 4.43(-01) &   ---     & 5.97(-01) & 5.98(-10)  &    ---     & 1.01(-09) \\
			& 20  & 2.60(+00) &   ---     & 7.05(+00) & 1.67(+00) &   ---     & 4.66(+00) & 5.71(-05)  &    ---     & 1.75(-04) \\
			& 30  & 2.64(+00) &   ---     & 8.11(+00) & 2.20(+00) &   ---     & 6.82(+00) & 2.64(-03)  &    ---     & 8.71(-03) \\
			&&&&&&&&&&\\
			$^{66}$Fe   & 1   & 2.49(+00) &   ---     & 2.49(+00) & 1.92(-01) &   ---     & 1.92(-01) & 8.71(-89)  &    ---     & 8.71(-89) \\
			& 3   & 2.51(+00) &   ---     & 2.51(+00) & 2.33(-01) &   ---     & 2.33(-01) & 5.85(-30)  &    ---     & 5.85(-30) \\
			& 5   & 2.54(+00) &   ---     & 2.54(+00) & 3.17(-01) &   ---     & 3.17(-01) & 7.89(-18)  &    ---     & 7.89(-18) \\
			& 10  & 4.55(+00) &   ---     & 4.55(+00) & 1.82(+00) &   ---     & 1.82(+00) & 2.19(-08)  &    ---     & 2.19(-08) \\
			& 20  & 2.41(+01) &   ---     & 2.41(+01) & 1.82(+01) &   ---     & 1.82(+01) & 1.75(-03)  &    ---     & 1.75(-03) \\
			& 30  & 3.47(+01) &   ---     & 3.47(+01) & 3.06(+01) &   ---     & 3.06(+01) & 6.98(-02)  &    ---     & 6.98(-02) \\
			&&&&&&&&&&\\
			$^{68}$Fe   & 1   & 5.04(+00) &   ---     & 3.16(+00) & 8.36(-01) &   ---     & 3.25(-01) & 7.57(-83)  &    ---     & 2.74(-83) \\
			& 3   & 5.06(+00) &   ---     & 3.18(+00) & 9.08(-01) &   ---     & 3.72(-01) & 2.91(-28)  &    ---     & 8.95(-29) \\
			& 5   & 5.09(+00) &   ---     & 3.23(+00) & 1.05(+00) &   ---     & 4.78(-01) & 9.55(-17)  &    ---     & 2.14(-17) \\
			& 10  & 8.59(+00) &   ---     & 4.69(+00) & 3.87(+00) &   ---     & 1.73(+00) & 1.02(-07)  &    ---     & 1.73(-08) \\
			& 20  & 5.04(+01) &   ---     & 9.46(+00) & 3.95(+01) &   ---     & 6.61(+00) & 5.40(-03)  &    ---     & 5.14(-04) \\
			& 30  & 7.46(+01) &   ---     & 1.07(+01) & 6.67(+01) &   ---     & 9.18(+00) & 1.87(-01)  &    ---     & 1.74(-02) \\
			&&&&&&&&&&\\
			$^{70}$Fe   & 1   & 1.44(+01) &   ---     & 9.95(+00) & 5.13(+00) &   ---     & 2.63(+00) & 7.03(-74)  &    ---     & 6.75(-74) \\
			& 3   & 1.44(+01) &   ---     & 9.98(+00) & 5.27(+00) &   ---     & 2.75(+00) & 1.24(-25)  &    ---     & 1.01(-25) \\
			& 5   & 1.46(+01) &   ---     & 1.01(+01) & 5.62(+00) &   ---     & 3.02(+00) & 1.78(-15)  &    ---     & 1.59(-15) \\
			& 10  & 1.91(+01) &   ---     & 1.39(+01) & 9.82(+00) &   ---     & 6.37(+00) & 1.62(-07)  &    ---     & 2.00(-07) \\
			& 20  & 2.82(+01) &   ---     & 2.62(+01) & 2.09(+01) &   ---     & 1.92(+01) & 1.96(-03)  &    ---     & 2.43(-03) \\
			& 30  & 3.08(+01) &   ---     & 2.90(+01) & 2.69(+01) &   ---     & 2.53(+01) & 5.78(-02)  &    ---     & 6.25(-02) \\
			&&&&&&&&&&\\
			\hline
	\end{tabular}}
\end{table*}


\newpage

\end{document}